\documentclass[traditabstract]{aa} 
\usepackage{graphics,txfonts,epsf,rotating,natbib, textcomp}
\pdfoutput=1
\bibpunct{(}{)}{;}{a}{}{,}

\def\muHz{\hbox{\textmu}Hz}

\begin{document}

\title{MOST observations of the roAp stars HD 9289, HD 99563, \\ and HD 134214\thanks{Based on data from the MOST satellite, a Canadian Space Agency mission, jointly operated by Dynacon Inc., the University of Toronto Institute for Aerospace Studies and the University of British Columbia, with the assistance of the University of Vienna.}}
\author{
		M. Gruberbauer\inst{1,2}
 		\and D. Huber\inst{1,3}
		\and R. Kuschnig\inst{1}
		\and W.~W. Weiss\inst{1}  
		\and D.~B. Guenther\inst{2}
		\and J.~M. Matthews\inst{4}
		\and A.~F.~J. Moffat\inst{5}
		\and J.~F. Rowe\inst{6}
		\and S.~M. Rucinski\inst{7}
		\and D. Sasselov\inst{8}
		\and M. Fischer\inst{9}
		}
\titlerunning{MOST observations of HD 9289, HD 99563, and HD 134214}
\authorrunning{M. Gruberbauer et al.}
\offprints{M. Gruberbauer : \\ \email{mgruberbauer@ap.smu.ca}}
\institute{
	University of Vienna, Institute for Astronomy (IfA), T\"urkenschanzstrasse 17, A-1180 Vienna, 
	\\ \email{werner.weiss@univie.ac.at, rainer.kuschnig@univie.ac.at}
	\and Department of Astronomy and Physics, Saint Mary's University, Halifax, NS B3H 3C3, Canada 
	\\ \email{mgruberbauer@ap.smu.ca, guenther@ap.smu.ca}		
	\and	Sydney Institute for Astronomy (SIfA), School of Physics, University of Sydney, NSW 2006, Australia
	\\ \email{dhuber@physics.usyd.edu.au}
	\and 	Department of Physics and Astronomy, University of British Columbia, 6224 Agricultural Road, Vancouver, BC V6T 1Z1, Canada
	\\ \email{matthews@astro.ubc.ca, gordonwa@uvic.ca}
	\and 	D\'{e}partment de physique, Universit\'{e} de Montr\'{e}al, C.P. 6128, Succ. Centre-Ville, Montr\'{e}al, QC H3C 3J7, Canada
	\\ \email{moffat@astro.umontreal.ca}		
	\and NASA Ames Research Center, Moffett Field, CA 94035, USA
	\\ \email{jasonfrowe@gmail.com}
	\and 	Department of Astronomy \& Physics, University of Toronto, 50 St. George Street, Toronto, ON M5S 3H4, Canada
	\\ \email{rucinski@astro.utoronto.ca}
	\and Harvard-Smithsonian Center for Astrophysics, 60 Garden Street, Cambridge, MA 02138, USA
	\\ \email{sasselov@cfa.harvard.edu}		
	\and Institute of Telecommunications, Vienna University of Technology, Gusshausstrasse 25/389, A-1040 Vienna, Austria
	\\ \email{michael.fischer@nt.tuwien.ac.at}}

\date{Received   / Accepted }
\abstract {We report on the analysis of high-precision space-based photometry of the roAp (rapidly oscillating Ap) stars \object{HD\,9289}, \object{HD\,99563}, and \object{HD\,134214}.
 All three stars were observed by the MOST satellite for more than 25 days, allowing unprecedented views of their pulsation. We find previously 
 unknown candidate frequencies in all three stars. We establish the rotation period of HD 9289 (8.5 d) for the first time and show that the star is pulsating in 
 two modes that show different mode geometries. We present a detailed analysis of HD\,99563's mode multiplet and find a new candidate frequency which appears independent of the previously known mode. Finally, we report on 11 detected pulsation frequencies in HD\,134214, 
 9 of which were never before detected in photometry, and 3 of which are completely new detections. Thanks to the unprecedentedly small frequency uncertainties, 
 the p-mode spectrum of HD\,134214 can be seen to have a well defined large frequency spacing similar to the well-studied roAp star \object{HD\,24712} (HR\,1217).}

\maketitle

\section{Introduction}  
\label{intro}
Rapidly oscillating Ap (roAp) stars are cool magnetic chemically peculiar stars with spectral types from early A to F0, luminosity class V. They were discovered by \citet{kurtz82} and pulsate with low amplitudes ($<$ 13\,mmag) at periods between 5 to 21 minutes. The oscillations are due to high-overtone ($n >15$) non-radial p-mode pulsations. The oblique pulsator model (hereafter OPM) as described by \citet{kurtz82} has been, to a large extent, a very applicable explanation of their observed amplitude variability. It predicts that the pulsation axis is aligned with the magnetic axis, which itself is oblique to the rotation axis of the star. The OPM has since been refined to include additional effects like the Coriolis force \citep{bigotdziem02}. For reviews on primarily observational aspects of roAp stars we refer to \citet{kurtz00} and \citet{kochukhov07}, and on theoretical aspects, to \citet{shiba03} and \citet{cunha05}.

The roAp stars offer great potential for asteroseismology. This potential is difficult to realise due to the small number of roAp stars known, their short pulsation periods and low amplitudes, and (apparently) very few observable modes. Recent investigations with the WIRE, MOST, and Kepler satellites, as well as groundbased high-resolution spectroscopy, have allowed us to study these objects with an unprecedented level of detail. 

Encouraged by the success of previous MOST observations of roAp stars \object{$\gamma$\,Equ} \citep{gruberbauer08a} and \object{10\,Aql} \citep{huber08b}, new observations of other roAp stars accessible to MOST were planned and performed with the primary goal of detecting additional frequencies. In this paper we report our findings on the targets HD\,9289, HD\,99563, and HD\,134214.

HD\,9289 ($V \approx 9.4$) was originally classified as ApScEu by \cite{bidelman73} and is now considered to be an A3p star \citep[see][and references therein]{ochsenbein81}. It was discovered to be a rapidly oscillating Ap star by \cite{kurtz94}.  The authors observed HD\,9289 on 22 nights, spanning several months from 1993 to 1994, and identified three individual frequencies around $\sim$ 1550 $\rm\mu Hz$. The latest results on this star were obtained spectroscopically by \cite{ryab07}, who were able to refine the values of some of its fundamental parameters. Most notably, they were able to estimate  $v\sin i = 10.5 \rm\,km\,s^{-1}$ and $\left<B_{\rm P}\right> = 2.0\,\rm kG$ for the first time. 

The first discovery of roAp pulsation in HD\,99563 was reported by \cite{dorokhov98}. The star was later studied extensively using long-timebase, ground-based photometry \citep{handler06}, as well as spectroscopy \citep{freyhammer09}. HD\,99563 is particularly interesting for the latter type of study because its pulsation shows large radial velocity amplitudes compared to other objects of this class. In all studies conducted so far it was found to pulsate in a single intrinsic pulsation mode which is split into a multiplet due to the effects predicted by the OPM. 

HD\,134214 was initially recognised to be a roAp star in 1985. A first thorough frequency analysis resulted in a single prominent mode with an amplitude of several mmag, at a frequency of 2949.6\,\muHz\ \citep{kreidl86}. No signs of rotational multiplets were detected. Subsequent observations did not reveal additional frequencies, but indicated that the frequency of the mode was periodically changing with a period of about 250 days \citep{kreidl94}. More than a decade later, time-resolved high-resolution spectroscopy of several roAp stars revealed signatures of previously unseen frequencies in almost all the stars in the sample, including HD\,134214 \citep{kurtz06}. This new variability was only observed in specific lines in the upper stellar atmosphere. Apparently, their amplitudes were too small for them to be detected with previous ground-based photometric observations. A 10-hour run with the MOST satellite could not confirm any of the newly reported signals, but it set an upper 2-$\sigma$ detection limit of 0.36 mmag for the corresponding amplitudes in the MOST bandpass \citep{cameron06}.
A later spectroscopic run dedicated to HD\,134214 \citep{kurtz07}, was intended to clarify the values of the newly seen frequencies. Due to the relatively short duration of the observations, and the apparent dependence of the frequency solution on the used spectral lines, no general list of new modes could be given. Nonetheless the general properties of the previously detected signals was confirmed and one of the additional frequencies detected in photometric observations. 

\section{Observations and data reduction}   			
\label{anal}

\begin{table}[t!]
	\caption{Summary of the MOST data sets of HD 9289, HD 99563, and HD 134214.}
	\centering
	\begin{tabular}{l l l l l l}
	\hline
	\hline
 	HD & $t_{\rm tot}$ & $N_{\rm tot}$ & $t_{\rm red}$ & $N_{\rm red}$ & $c_{\rm eff}$ \\
		 &  [d]  &  &  [d] & & \\ 
	\hline 
	HD 9289      & 26.27   & 41884  & 26.06    &  37635  &  0.632  \\
	HD 99563    & 27.53   & 62261  & 26.68    &  55176  &  0.773  \\
	HD 134214 &  25.96   &  35099  & 23.92   &  31244  &  0.459  \\
	\hline
	\hline
	\end{tabular}
	\label{tab:redres}
	\tablefoot{
	$t_{\rm tot}$,  $N_{\rm tot}$: total timebase and number of data points for MOST observing run; $t_{\rm red}$, $N_{\rm red}$: timebase and number of data points for final light curve; $c_{\rm eff}$: effective duty cycle for final light curve (see text for a definition)}
\end{table}%

  MOST (Microvariability and Oscillation of STars) is a Canadian microsatellite designed for the rapid-cadence observation of stellar photometric variability on a level of a few parts-per-million (ppm). The satellite is in a sun-synchronous polar orbit, above the terminator, and has an orbital period of 101 min. It is able to observe bright stars in its continuous viewing zone (CVZ) for up to 2 months. It houses a Maksutov telescope with an aperture of 15 cm, a custom broadband filter (3500 -- 7000\,\AA), and two CCDs. One CCD was originally designed to be exclusively employed for satellite guiding (Tracking CCD), while the other CCD should collect the scientific observations (Science CCD). To accomplish its task more accurately, the latter is equipped with a Fabry-lens array covering a small part of the detector area. When observing stars in the upper brightness range of the instrument, the Fabry lenses are used to reduce the jitter of the image that is caused by pointing instabilities and to avoid saturation by spreading the light over a very large number of pixels. In contrast, dimmer programme stars and guide stars are observed in the open field (Direct Imaging - DI). A more detailed description of the early mission is given by \cite{walker03}. 
In 2006 the Tracking CCD stopped working, and the Science-CCD software was modified to additionally provide the information for the attitude control system (ACS). Since the ACS constantly needs information about the satellite pointing, stars are observed with short exposure times, but the individual exposures are stacked to reach and improve upon the original levels of photometric precision achieved by MOST. 

As explained by \cite{reegen06}, MOST suffers varying amounts of stray light as it orbits about the partially illuminated Earth, which causes artificial signal at multiples of the MOST orbital frequency. In addition, this non-stellar signal is modulated throughout an observational run because of changes in the average albedo of the illuminated limb of the Earth as seen from the position of the satellite. This modulation occurs on timescales comparable to a sidereal day. Longer-term instrumental drifts are sometimes also present, but MOST has shown instrumental stability of about 1 mmag over intervals of years. The MOST team has become very familiar with these effects, and photometric reduction pipelines for the different MOST observing modes \citep{reegen06,rowe06, huber08a, hareter08} effectively filter the artifacts in many cases while leaving the stellar signals intact.


All three roAp stars discussed in this paper have been observed using the DI mode. The general properties of these data sets are summarised in Table\,\ref{tab:redres}. This table also includes what we call the ``effective duty cycle"
\begin{equation}
c_{\rm eff} = \frac{\left<\Delta t\right> \left ( N-1 \right )}{t_{\rm obs}},
\end{equation}  
where $\left<\Delta t\right>$ is the median time stamp difference between two adjacent data points, $N$ is the total number of observations, and $t_{\rm obs} = t \left ( N \right ) - t \left (0 \right )$ is the total time baseline of the resulting time series. The median is not very sensitive to outliers, and so closely reflects the effective sampling of the time series, even if sporadic under-sampling occurs due to larger gaps. 

One can also define an effective sampling frequency
\begin{equation}
\nu_{\rm eff} = \frac{1}{2 \left<\Delta t\right>},
\end{equation}
similar to the definition of the Nyquist frequency. This, in combination with the window function, is a good measure of the actual sampling quality of the data.

\subsection{HD\,9289}
\label{sec:obs9289}
\begin{figure*}[t]	
\centering
	\includegraphics[width=0.75\textwidth]{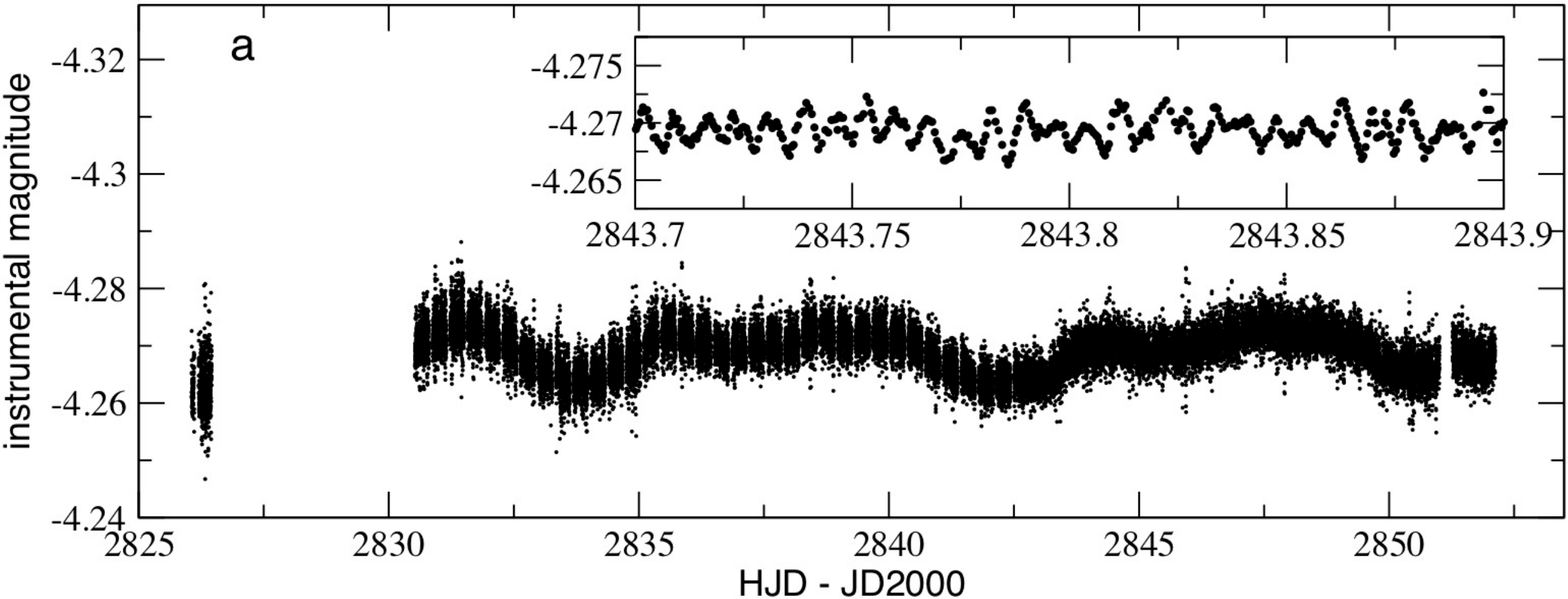}
	\includegraphics[width=0.75\textwidth]{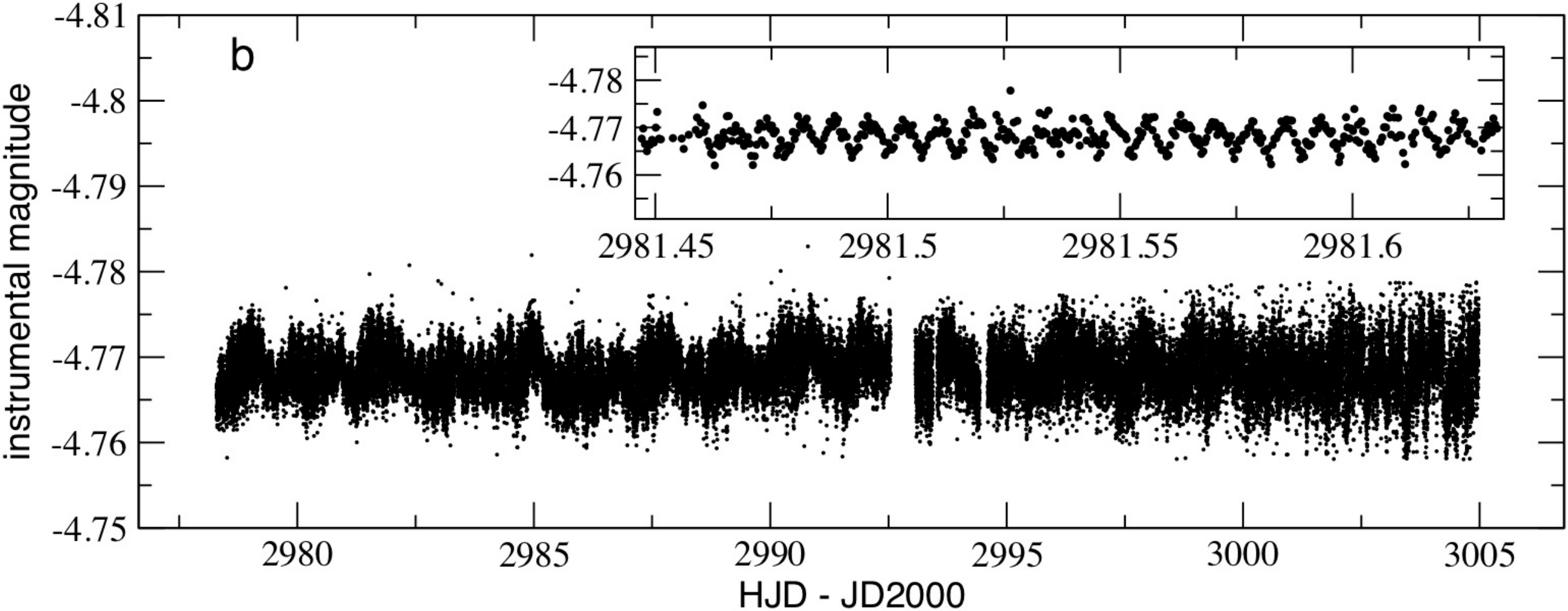}
	\includegraphics[width=0.75\textwidth]{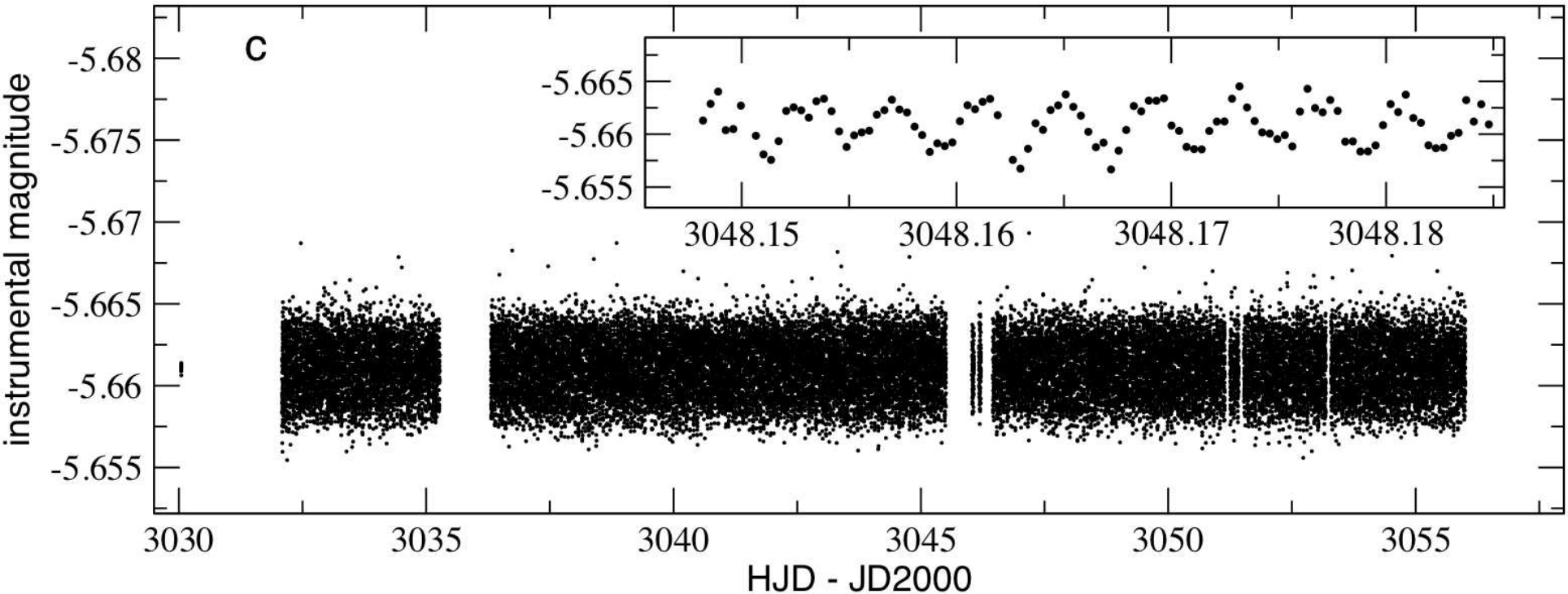}
	\caption{Reduced MOST light curves of (a) HD\,9289, (b) HD\,99563, and (c) HD\,134214. The inserts shows a small portion of a version of the light curve where any variability not due to roAp pulsation has been subtracted. For the insert in panel (a), a 5-point running average is displayed instead of the unbinned data, in order to enhance the clarity of the weaker pulsation signature. Note that JD2000 = 2451545.0}
	\label{fig:redlc}
\end{figure*}

MOST observed HD\,9289 for about 22 of the 26 days during September 27 to October 23, 2007. After the first day of observation a 4-day gap occurred, and during the first 16.4 days of observation MOST alternated between two fields of view. During this setup HD\,9289 was monitored for 4 consecutive orbits, after which MOST switched to the other primary target for the duration of 1 orbit, and then back to HD\,9289. For the remaining 10.6 days, however, HD\,9289 was observed without interruption. The vast majority ($ > 99\,\%$) of the measurements consisted of image stacks of 20 exposures each, with a total exposure time of 32.13 seconds. 

The data were reduced with the DI-pipeline developed by \cite{huber08a}. Different pipeline parameters were set to produce two versions of the light curve: the ``final light curve" version, where instrumental effects are suppressed as far as possible, and the ``consistency check" version that underwent a less invasive treatment and more closely reflects the raw data. For each version of the reduction, about 4200 frames (about 10\% of the total number in the time series) were rejected during the initial stages of the reduction process because of heavy stray light pollution, and/or excessive cosmic-ray incidence in the proximity of the Southern Atlantic Anomaly (SAA). The values cited in Table\,\ref{tab:redres} refer to the light curve with the smallest instrumental contributions. Apart from increased scatter at high-stray light orbit phases and possible long-term trends, no obvious instrumental effects are present in the final light curve (see Fig.\,\ref{fig:redlc}). It shows a regular double-hump variation with a period of about 8.5 days which has been tracked for almost 3 cycles. This is actually rotational modulation of the mean stellar brightness, probably due to spots. It is the first time that such a signature has been detected for HD\,9289. The insert in panel (a) of Fig.\,\ref{fig:redlc} also demonstrates the presence of the short-period roAp pulsations.

\subsection{HD\,99563}
\label{sec:obs99563}

\begin{figure}[t]	
\centering
	\includegraphics[width=\columnwidth]{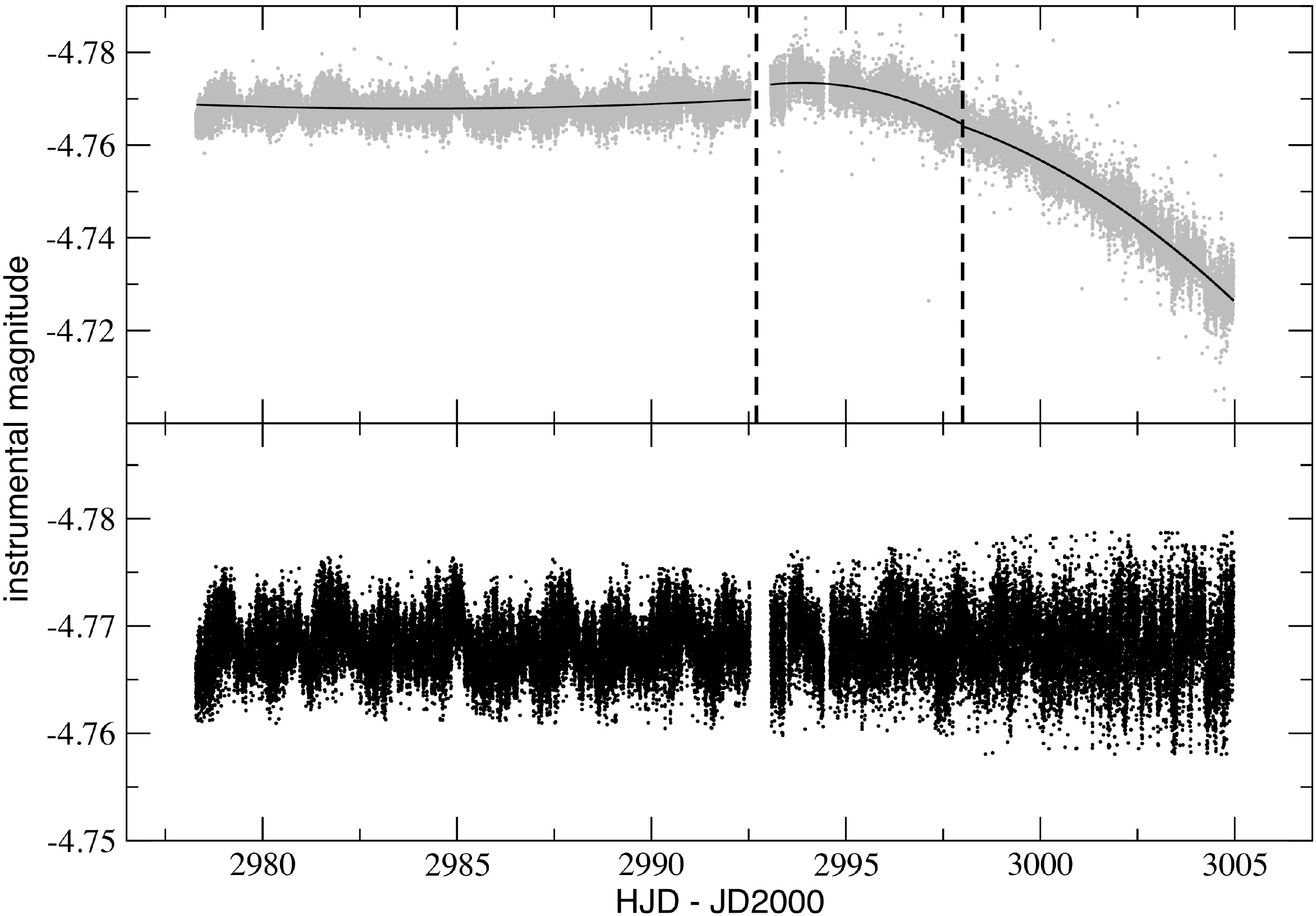}
	\caption{Light curve of HD\,99563 after processing with the reduction pipeline (top) and after subtraction the instrumental long-term trends (bottom).
	The dashed vertical lines separate the individually fitted subsets of the time series.}
	\label{fig:hd99563detrend}
\end{figure}

MOST monitored HD\,99563 for about 26 days during February 25 to March 23, 2008. However, the first day of observation did not yield any usable data. A second target was observed in a different field, hence MOST was re-oriented every 3rd orbit for about 50 minutes, decreasing the duty cycle. Almost all of the remaining frames consist of 15 stacked exposures, with a total integration time of 32.32 seconds. With respect to the reduction, the same steps were performed as for HD\,9289. However, due to a change in the observation setup at $\rm JD - JD2000 \approx 2992.5$ (JD2000 = 2451545.0) there is a half-day gap in the time series. The data obtained beyond this gap are of slightly lower quality due to an increase in stray light, and a long-term instrumental trend, which is also reflected in the board temperature telemetry\footnote{The CCD focal plane temperature is kept constant to within 0.1~K, but the rest of the satellite temperature can vary by about 2 K during each MOST orbit. This effect can be more pronounced if MOST is switching between two different, widely separated fields during each orbit}. To improve the situation the latter parts of the light curve were treated separately using different parameters for the reduction pipeline. The data set was then homogenised by fitting and subtracting polynomials of second order to the two parts (see Fig.\,\ref{fig:hd99563detrend}). The final light curve is shown in Fig.\,\ref{fig:redlc}b. There is obvious variability on timescales much longer than what is expected from the pulsation alone. Some of the variability is caused by the OPM effect, which modulates the pulsation amplitude over one stellar rotation cycle. A second variability component is due to rotational modulation of the mean stellar brightness, which has already been detected in previous studies \citep[e.g.,][]{handler06}. Because of the broad MOST bandpass (about 350 - 700 nm) and the wavelength dependence of Ap rotational modulation, this signal is much weaker than what Handler et al. found in the ground-based data with a standard blue filter. It is also less pronounced than the spot effects in HD\,9289. A third component is purely instrumental and caused by complex onboard temperature variations due to the target switching. We did not attempt to remove this component during the reduction process in order to not distort the intrinsic rotational effects on the pulsation.

\subsection{HD\,134214}
\label{sec:obs134214}

\begin{figure}[t]	
\centering
	\includegraphics[width=\columnwidth]{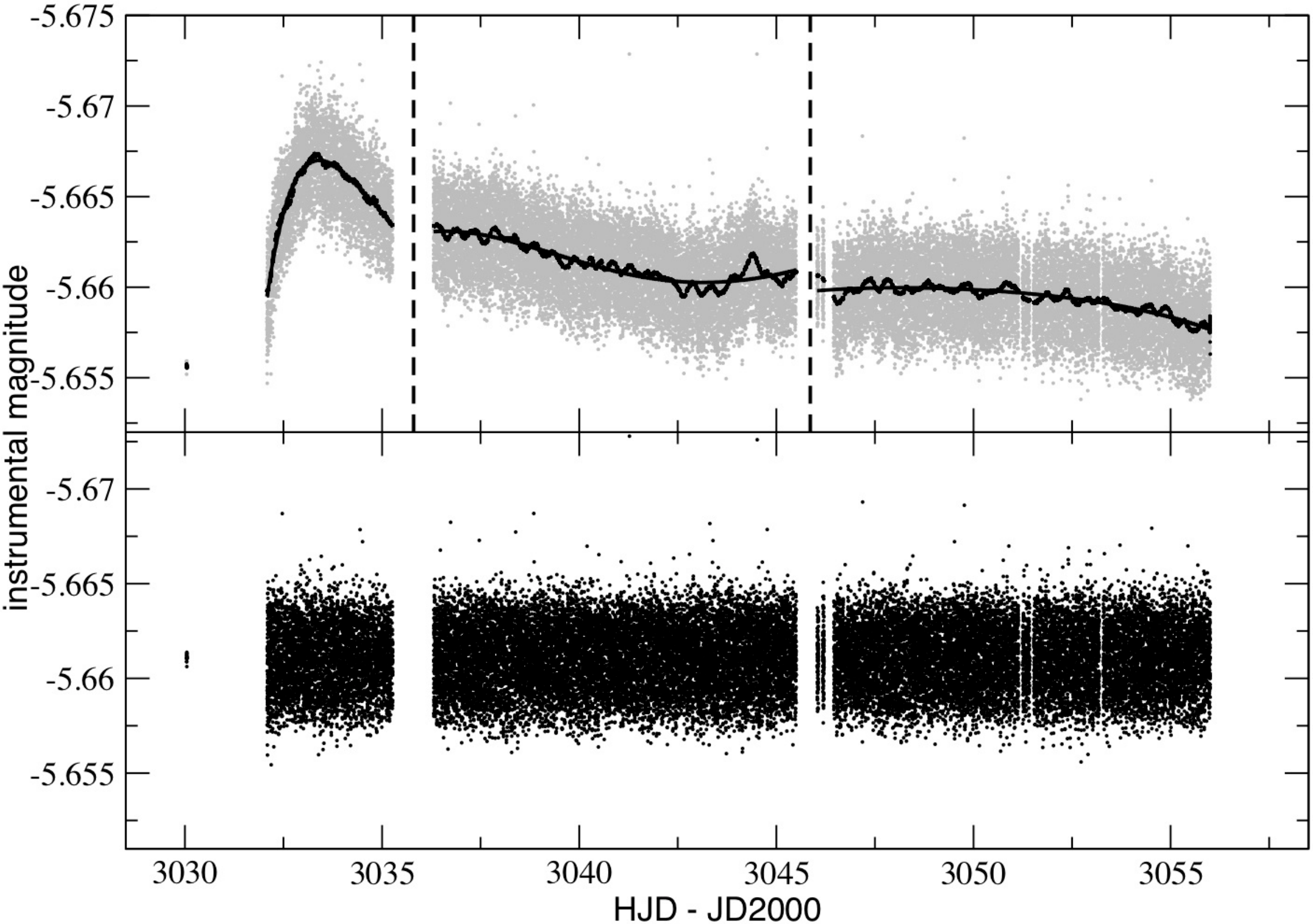}
	\caption{Same as Fig.\,\ref{fig:hd99563detrend} but for HD\,134214. On top of the long-subset polynomial fits the top panel also shows a 200 pt running average used as a high-pass filter.}
	\label{fig:hd134214detrend}
\end{figure}

HD\,134214 was observed for about 26 days during April 18 to May 14 2008. Each MOST orbit was shared equally between this star and another science target field throughout the run, reducing the duty cycle to about 50 per cent. The majority of images are stacks of 19 exposures (total integration time: 30.2 seconds). Again, we used the same steps to reduce the data as in the analysis of HD\,9289, yielding two versions of the light curve. The only difference is that, after reduction, long-term trends, which correlate with board temperature variation, are still visible in the data. These have been removed from the light curve (see Fig.\,\ref{fig:hd134214detrend}) in two ways. First we fitted and subtracted fourth order polynomials from three distinct subsets of the light curve clearly separated by larger gaps. However, we also noticed a modulation of the board temperature on short time scales due to the target switching. Due to the absence of any obvious rotational signature, we therefore also used a high-pass filter, removing any variations on time-scales longer than the MOST orbital period. The latter approach produced fewer instrumental artefacts, and the result is referred to as the ``final light curve" for HD\,134214. However, we repeated all further analysis steps also for the less intrusive approach to make sure that we did not actually introduce any fake signal. The final light curve is dominated by the roAp pulsation and does not show any significant long-term variability after removing the instrumental trends.

\section{Frequency analysis}   	
\label{sec:freqana}		
The final light curves were analysed using the latest version of SigSpec \citep{reegen07}. SigSpec 2.0 performs an iterative multi-cosine fit to time-series data, based on the ``spectral significance" (or sig) as a significance criterion. The sig is defined as the negative logarithm of the probability that a periodic signal of the form $a \cos\left ( 2 \pi \nu t - \theta \right)$ (where $a$ is the amplitude, $\nu$ is the frequency, and $\theta$ is usually called the phase) is caused by white noise. SigSpec starts by computing the sig for a given range in frequency and finding the frequency, amplitude, and phase that correspond to the smallest false-alarm probability.  It then refines the solution by performing a least-squares fit in the time domain. Subsequently, the solution is subtracted from the time-series, and the sig is computed for the residuals. This continues iteratively, each time adding another component to the least squares multi-cosine fit, until no signal is found with a sig above a certain threshold. Formally, an amplitude-to-noise ratio of 4, which is generally assumed to be a conservative lower limit, corresponds to $\rm sig \approx 5.46$. Self-evidently, the definition of the sig is only truly valid in the case of a single sinusoidal signal in the presence of white noise. The results therefore have to be interpreted with care, and often a higher threshold than 5.46 is sensible.

Even when data reduction is performed thoroughly, a number of significant instrumental signals due to stray light or target field switches can remain in MOST time series. Any gaps due to target switching and/or outlier removal during high stray light phases make it imperative to correctly pre-whiten these signals together with the intrinsic frequencies. To properly take this into account in the analysis of the three roAp stars we proceeded as follows. 

In the first step, SigSpec was used to calculate the most probable multi-cosine solution down to a sig threshold of 5.46 for each reduced light curve. The upper limit in frequency for the SigSpec analysis was set for each star so we could detect a possible harmonic to the main pulsation frequency. The same calculation was performed for time series composed of the mean background measurements, and also board temperature measurements, as provided by the reduction pipeline. For each star, the resulting frequencies of all time series were then compared and put into one of three categories. 
\begin{enumerate}
\item \textbf{Instrumental artefacts}. (a) Identifiable in both the stellar photometry and in either background or temperature telemetry, and/or (b) obviously related to a harmonic of MOST orbital frequency ($\approx14.19 \rm\,d^{-1}$). Their amplitudes vary drastically in the stellar time series when different pipeline parameters are used. 
\item \textbf{Intrinsic low-frequency variability}. Any highly significant signals at lower frequencies (i.e., due to rotational modulation caused by spots) and found in both reduced light curves but not in the background/telemetry.
\item \textbf{Intrinsic pulsation (candidates)}. Possible pulsation frequencies, in particular signals in the range associated with roAp oscillations, that are found in the more and the less invasively reduced light curves but not in the background/telemetry. 
\end{enumerate}

Frequencies in categories 2 and 3 were checked for consistency with Period04 \citep{lenz05}. For further time-resolved analysis of the pulsation candidates, all signals except those in category 3 were subtracted from the light curves. The uncertainties for frequencies, amplitudes, and phases are analytical lower limits calculated with Period04. We cross-checked these values with the SigSpec results using the heuristic relations from \cite{kallinger08}. The latter provide a reliable upper limit in the uncertainties. Interpreting these as $4\sigma$ confidence limits we find both results to be highly consistent.

\subsection{HD\,9289}
The results of the frequency analysis for HD\,9289 are presented in Table \,\ref{tab:freqana9289}. The MOST data reveal the presence of 6 intrinsic significant frequencies between 1570 and 1590 \muHz. In the low-frequency range, we can also find signals corresponding to the rotational modulation. However, in contrast to the high frequency signals, the exact values of these low frequencies depend on the reduction procedure used. This reflects the fact that it is very difficult to preserve the exact nature of long-term trends in space photometry due to the lack of calibration sources and procedures. This can lead to problems with analyses using the Discrete Fourier transform (DFT), which expects a periodic harmonic behaviour of the variations that are analysed. 

Since we can assume that the rotational modulation of roAp stars, caused by spots, is stable across many cycles, any deviations are likely instrumental. We can therefore apply the phase dispersion minimization technique \citep[PDM,][]{stellingwerf78} to obtain an estimate of the rotation frequency based on the periodicity of the signal, more independent of the exact preservation of its shape. We used the PDM technique on the two reduced light curve versions, as well as on a raw data version that was filtered as to best preserve the intrinsic long term variations. We then calculated the mean and standard deviation of the resulting rotation frequencies, noted as $\nu_{\rm rot, PDM}$ in Table\,\ref{tab:freqana9289}. Luckily, this crude estimate can be improved upon, since there are other ways to get a less reduction-dependent estimate of the rotation frequency, $\nu_{\rm rot, puls}$, as discussed in Sect.\,\ref{sec:discuss9289}. 

\begin{figure}[t]	
\centering
	\includegraphics[width=\columnwidth]{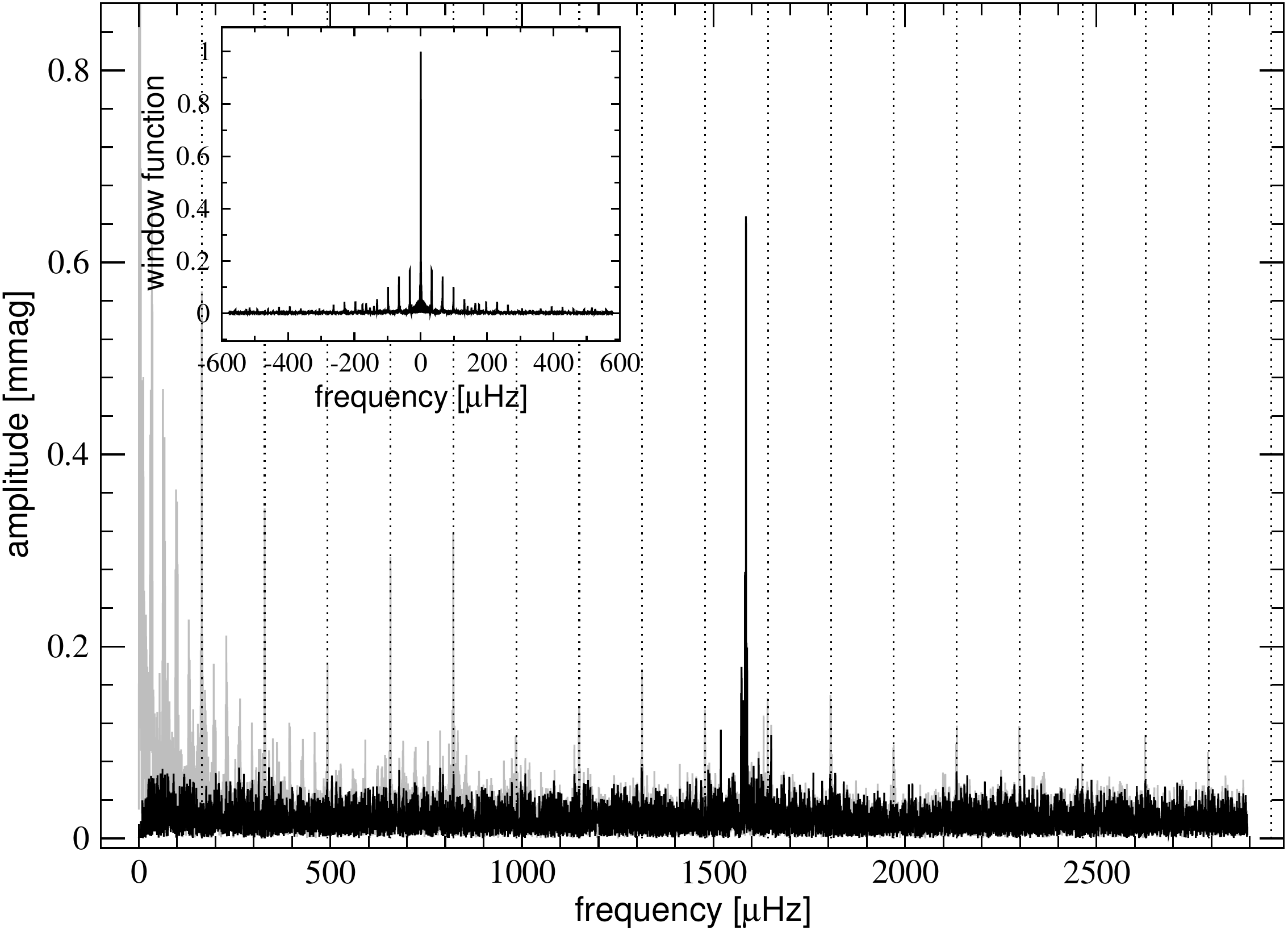}
	\caption{DFT of the HD\,9289 data without any pre-whitening (grey) and after pre-whitening of all category 1 and 2 signals as defined in Sect.\,\ref{sec:freqana} (black). The insert shows the window function of the data set. Vertical dotted lines show the MOST orbital frequency and its integer mulitples.}
	\label{fig:hd9289_ampspec}
\end{figure}

A large number of additional weak signals with frequencies below $10\rm\,d^{-1}$ are present as well, but a comparison with the background and board temperature reveals that most of them are instrumental. We conclude that we cannot reliably assess the intrinsic nature of any additional low-frequency variability. Fig.\,\ref{fig:hd9289_ampspec} compares the amplitude spectrum before and after pre-whitening all signals that are members of category 1 and 2. Fig.\,\ref{fig:hd9xcombo} shows a schematic spectrum of the pulsation multiplets.

\begin{table}[t!]
	\caption{Frequency analysis results for HD\,9289.}
	\centering
	\begin{tabular}{l r r r r r r}
	\hline
	\hline
 	id & frequency & amplitude & $\theta$ & sig & SNR \\
		 &  [\muHz]  & [mmag] &  [rad]  & &  \\ 
	\hline 
	$\nu_1$ & 1584.936(7) & 0.63(2) & 5.05(2) & 259.8 & 28.6 \\
	$\nu_2$ & 1582.22(2) & 0.24(2) & 4.68(6) & 48.4 & 10.8 \\
	$\nu_3$ & 1587.67(2) & 0.16(2) & 5.8(1) &  20.4 & 7.1 \\
	$\nu_4$ & 1573.27(3) & 0.15(2) & 6.0(1) & 23.2 & 6.5 \\
        $\nu_5$ & 1571.92(3) & 0.14(2) & 1.6(1) & 16.4 & 6.1 \\
        $\nu_6$ & 1577.33(4) & 0.11(2) & 5.7(1) & 10.5 & 4.9 \\
        $\nu_{\rm rot, PDM}$ & 1.35(1) & & \\
        $\nu_{\rm rot, puls}$ & 1.361(7) & & \\

	\hline
	\hline
	\end{tabular}
	\label{tab:freqana9289}
	\tablefoot{
	Phase $\theta$ is given according to the definition $\cos\left(2\pi\nu t - \theta\right)$ with respect to the epoch HJD 2454371.05691. The numbers in brackets are the errors in units of the last digit. SNR stands for signal-to-noise ratio.}
\end{table}

\subsection{HD\,99563}
\label{sec:hd99563_freqana}

\begin{figure}[t]	
\centering
	\includegraphics[width=\columnwidth]{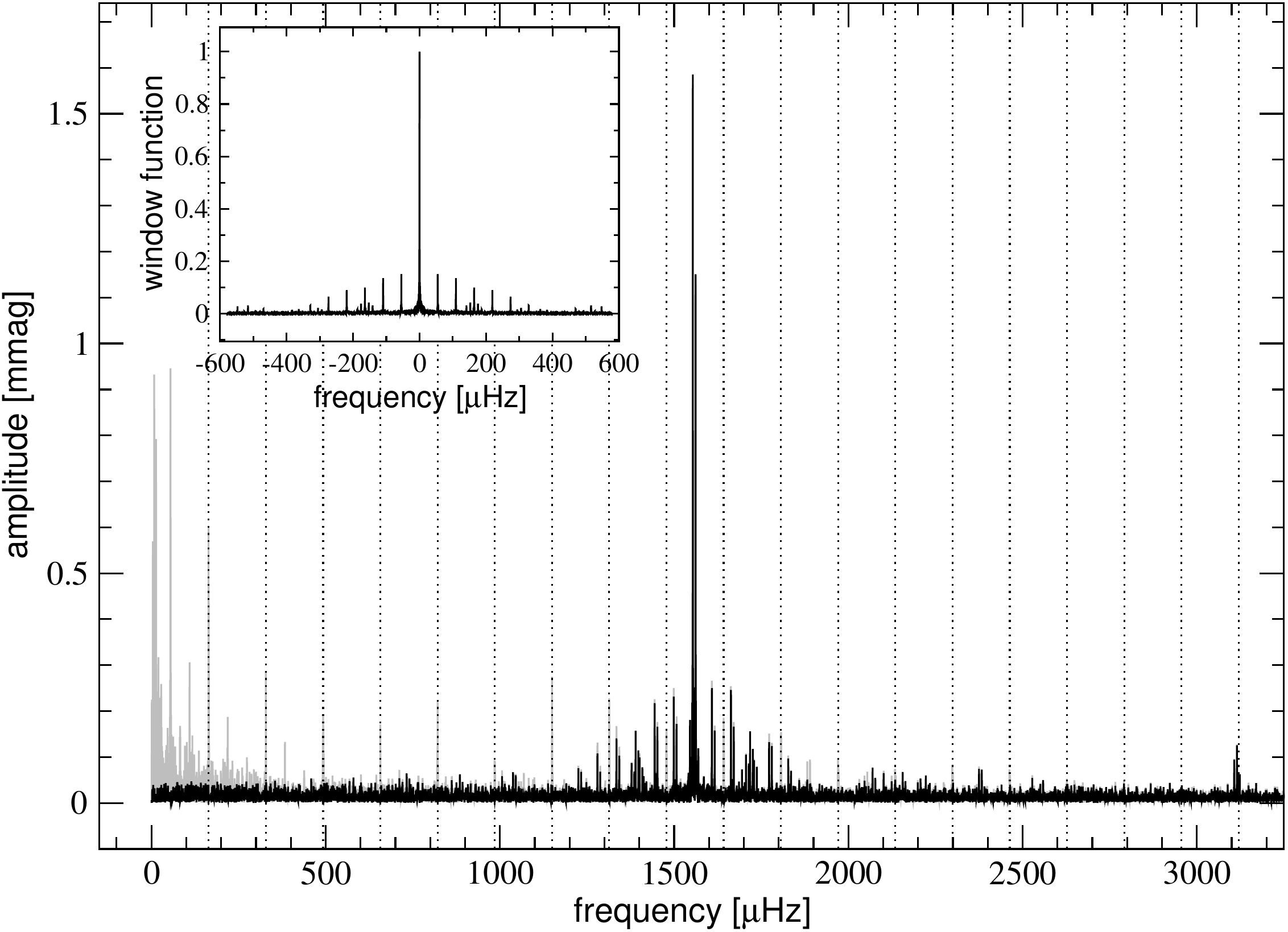}
	\caption{Same as Fig.\,\ref{fig:hd9289_ampspec} but for HD\,99563. Notice the presence of a small signal ``hump" due to the harmonic to the main pulsation frequency, and combination frequencies, at about 3110 \muHz. Another small hump at about 2380 \muHz\, is caused by aliasing.}
	\label{fig:hd99563_ampspec}
\end{figure}

As presented in Fig.\,\ref{fig:hd99563_ampspec}, a large number of instrumental artefacts with substantial amplitudes can be identified in the low-frequency regime. This stems from target switching, as discussed in Sect.\,\ref{sec:obs99563}. Nonetheless, we can identify what seems to be the stellar rotation frequency and one of its harmonics. However, the value found for the ``fundamental" is  3.902(5)\,\muHz, which is inconsistent with the precise value given by \cite{handler06}. The second frequency interpreted to be the first harmonic found at 7.970(3)\,\muHz\, is in better agreement with the previously published value, but it is still too far off given the formal uncertainties, and incompatible with our value of the ``fundamental".  A PDM analysis of the least noisy parts of the data yields a value of 3.95\,\muHz, but the result is not well constrained due to the small amplitude of the variation. We thus conclude that we cannot improve the accuracy of the rotation period using the mean light variation.

Concerning pulsation frequency candidates, we treated HD\,99563 in a special way, since its main (and hitherto only) pulsation frequency is already so well studied. We first performed our unbiased, unconditioned analysis as explained at the beginning of the section. The results for this approach are presented in Table\,\ref{tab:freqana99563}. As expected, we find 7 more or less equally spaced frequencies centred at $\approx 1557.7$ \muHz. We also find a second group of frequencies composed of integer multiples (commonly called ``harmonics") and linear combinations of the septuplet frequencies. The broad picture is mostly consistent with previous observations. We do detect the (-2,+2) components that \citeauthor{handler06} could not find in their ground-based data. However, we cannot find the outermost (-4,+4) nonuplet components that \citeauthor{freyhammer09} report. A closer look reveals additional discrepancies. The frequency value for the central component of the septuplet differs by almost $3\sigma$. We also find a number of additional significant frequencies in between the septuplet. These candidates are labeled $\nu_8$ to $\nu_{12}$ in Table\,\ref{tab:freqana99563}. 

Most of the remaining frequencies found are consistent with well-known stray light artefacts, situated at multiples of the MOST orbital frequency. What is left can be explained through another effect: modulation of the {\it stellar} signal with the MOST orbital frequency $\nu_{\rm MOST}$. We find formally significant frequencies at $\nu_{2} \pm n \nu_{\rm MOST}$ and $\nu_{3} \pm n \nu_{\rm MOST}$, where $n$ is an integer up to $n=3$. This is not aliasing, as one might think when looking at the window function of the data set, but rather an actual instrumental modulation of the measured stellar signal. These effects cannot be easily corrected for, but they can easily be identified in the course of the frequency analysis.

The rotation period of HD\,99563 can't be determined to high accuracy from the light modulation signal in the MOST time series. But it can be obtained independently, and in an unbiased way, from the average spacing of the OPM multiplet. We find $\left< \nu_{\rm rot} \right> = 3.985(7)$\,\muHz, which is compatible with the literature, but obviously not precise enough to be considered an improvement.

We also chose a second approach that would more closely reflect our prior knowledge of HD\,99563's pulsation and rotation. \citeauthor{handler06} give a quite precise value for the rotation frequency of the star $\nu_{\rm rot, Handler} = 3.9749(1)$\,\muHz. Using this value, we forced a fit to a nonuplet with the corresponding spacing, looking for whatever frequency gave the best results. The fitting procedure was carried out using Period04, after all category 1 and 2 signals had already been subtracted. The result is $\nu_{\rm p} = 1557.67(1)$\,\muHz, which is closer to what \citeauthor{handler06} found for their data set. If this value is changed to the exact previously known value, the residuals to the fit show no significant change. Therefore, we also fixed the pulsation frequency to $\nu_{\rm p} = 1557.6529$\,\muHz. We then proceeded with a pre-whitening sequence to look for additional frequencies. After each iteration we allowed the multi-sine fit to improve the frequencies, amplitudes, and phases for any additional signal found, as well as amplitudes and phases of all nonuplet frequencies. I.e., we kept $\nu_{\rm p}$ and $\nu_{\rm rot}$ fixed to the values matching the results of \citeauthor{handler06}. The outcome of this analysis is presented in Table\,\ref{tab:forcedfreqana99563} and a comparison of the forced and the unconditional solutions is shown Fig.\,\ref{fig:hd9xcombo}.

\begin{table}[t!]
	\caption{Unconditional multi-frequency fit results for HD\,99563.}	
	\centering
	\begin{tabular}{l r r r r r r}
	\hline
	\hline
 	id & frequency & amplitude & $\theta$ & sig & SNR \\
		 &  [\muHz]  & [mmag] &  [rad]  & &  \\ 
	\hline 
	$\nu_1 = \nu_{\rm p}$ & 1557.68(1) & 0.23(1) & 4.58(4) & 88.1 & 15.8 \\ 
	$\nu_2 \approx \nu_{\rm p - 1}$ & 1553.683(2) & 1.50(1) & 2.634(7) & 1969.5 & 102.5 \\ 
	$\nu_3 \approx \nu_{\rm p + 1}$ & 1561.657(3) & 1.11(1) & 4.796(9) & 1184.8 & 73.3\\ 
	$\nu_4 \approx \nu_{\rm p - 2}$ & 1549.68(4) & 0.06(1) & 2.8(2) & 6.7 & 4.2 \\ 
	$\nu_5 \approx \nu_{\rm p + 2}$ & 1565.67(3) & 0.08(1) & 0.6(1) & 11.9 & 5.3 \\ 
	$\nu_6 \approx \nu_{\rm p - 3}$ & 1545.76(2) & 0.16(1) & 1.08(6) & 44.0 & 10.8 \\ 
	$\nu_7 \approx \nu_{\rm p + 3}$ & 1569.56(3) & 0.10(1) & 0.4(1) & 18.5 & 6.5 \\ 
	$\nu_{8}$ & 1553.21(3) & 0.12(1) & 2.5(1) & 10.9 & 8.3 & \\ 
	$\nu_{9}$ & 1561.83(4) & 0.10(1) & 1.0(1) & 7.7 & 6.9 \\ 
	$\nu_{10}$ & 1552.35(4) & 0.06(1) & 2.5(2) & 8.3 & 4.2 \\ 
	$\nu_{11}$ & 1554.01(4) & 0.17(2) & 5.42(5) & 6.4 & 12.5 \\ 
	$\nu_{12} = \nu_{\rm cand}$ & 1555.84(4) & 0.06(1) & 0.6(2) & 7.1 & 4.2 \\ 
	
	$2\nu_{\rm p}$ & 3115.36(2) & 0.13(1) & 2.78(8) & 32.7 & 9.7 \\ 
	$2\nu_{\rm p - 1}$ & 3107.46(3) & 0.09(1) & 0.3(1) & 18.1 & 7.1 \\ 
	$2\nu_{\rm p} - \nu_{\rm rot}$ & 3119.38(4) & 0.06(1) & 2.8(2) & 9.5 & 4.8 \\ 
	$2\nu_{\rm p + 1}$ & 3123.30(4) & 0.06(1) & 4.8(2) & 7.7 & 4.3 \\ 
	
	$\nu_{\rm rot}$ & 3.985(7) & & & & & \\
	\hline
	\hline
	\end{tabular}
	\label{tab:freqana99563}
	\tablefoot{
	Phase $\theta$ is given according to the definition $\cos\left(2\pi\nu t - \theta\right)$ with respect to the epoch HJD\,2454523.29297. $\nu_{\rm rot}$ was derived from a weighted average of the spacings of the main septuplet.}
\end{table}

\begin{table}[t!]
	\caption{Multi-frequency fit results for HD\,99563, forcing the frequency solution to match $\nu_{\rm rot}$ given in \cite{handler06}.}
	\centering
	\begin{tabular}{l r r r r r r}
	\hline
	\hline
 	id & frequency & amplitude & $\theta$ & SNR \\
		 &  [\muHz]  & [mmag] &  [rad]  & &  \\ 
	\hline 
	$\nu_1 \equiv \nu_{\rm p}$ & 1557.6529 & 0.23(1) & 4.35(4) & 14.9 \\ 			
	$\nu_2 \equiv \nu_{\rm p - 1}$ & 1553.678(2) & 1.62(1) & 2.619(6) & 105.1  \\ 
	$\nu_3 \equiv \nu_{\rm p + 1}$ & 1561.628(6) & 0.41(1) & 3.73(2) & 26.1  \\ 
	$\nu_4 \equiv \nu_{\rm p - 2}$ & 1549.70(4) & 0.06(1) & 3.0(2) & 3.9  \\ 
	$\nu_5 \equiv \nu_{\rm p + 2}$ & 1565.60(3) & 0.07(1) & 0.1(1) & 4.7  \\ 
	$\nu_6 \equiv \nu_{\rm p - 3}$ & 1545.73(2) & 0.15(1) & 0.87(6) & 10.1  \\ 
	$\nu_7 \equiv \nu_{\rm p + 3}$ & 1569.58(2) & 0.10(1) & 0.6(1) & 6.1  \\ 
	$\nu_8 \equiv \nu_{\rm p - 4}$ & 1541.8(1) & 0.02(1) & 4.8(6) & 1.1 \\ 
	$\nu_9 \equiv \nu_{\rm p + 4}$ & 1573.6(2) & 0.01(1) & 3.7(8) & 0.8 \\ 
	$\nu_{10}$ & 1561.671(3) & 0.94(1) & 5.32(1) & 59.5  \\ 
	$\nu_{11}$ & 1553.38(1) & 0.17(1) & 3.80(6) & 11.2  \\ 
	$\nu_{12}$ & 1554.08(1) & 0.14(1) & 5.95(6) & 9.4  \\ 
	$\nu_{13} = \nu_{\rm cand}$ & 1555.85(4) & 0.06(1) & 0.7(2) & 3.9  \\ 
	
        $2\nu_{\rm p}$ & 3115.36(2) & 0.13(1) & 2.76(8) & 9.3  \\ 
        $2\nu_{\rm p - 1}$ & 3107.46(3) & 0.09(1) & 0.4(1) & 6.9 \\
        $2\nu_{\rm p} - \nu_{\rm rot}$ & 3119.38(4) & 0.06(1) & 2.8(2) & 4.8  \\ 
        $2\nu_{\rm p + 1}$ & 3123.29(4) & 0.06(1) & 4.8(2) & 4.3  \\ 	
        $\nu_{\rm rot}$ & 3.9749 & & & & & \\
	\hline
	\hline
	\end{tabular}
	\label{tab:forcedfreqana99563}
	\tablefoot{
	Frequencies which are not part of the multiplet or harmonics have been labelled in order of decreasing SNR. The labels do not necessarily correspond to the frequencies in Table\,\ref{tab:freqana99563}. Phase $\theta$ is given according to the definition $\cos\left(2\pi\nu t - \theta\right)$ with respect to the epoch HJD\,2454523.29297.}
\end{table}

\subsection{HD\,134214}
\begin{figure}[t]	
\centering
	\includegraphics[width=\columnwidth]{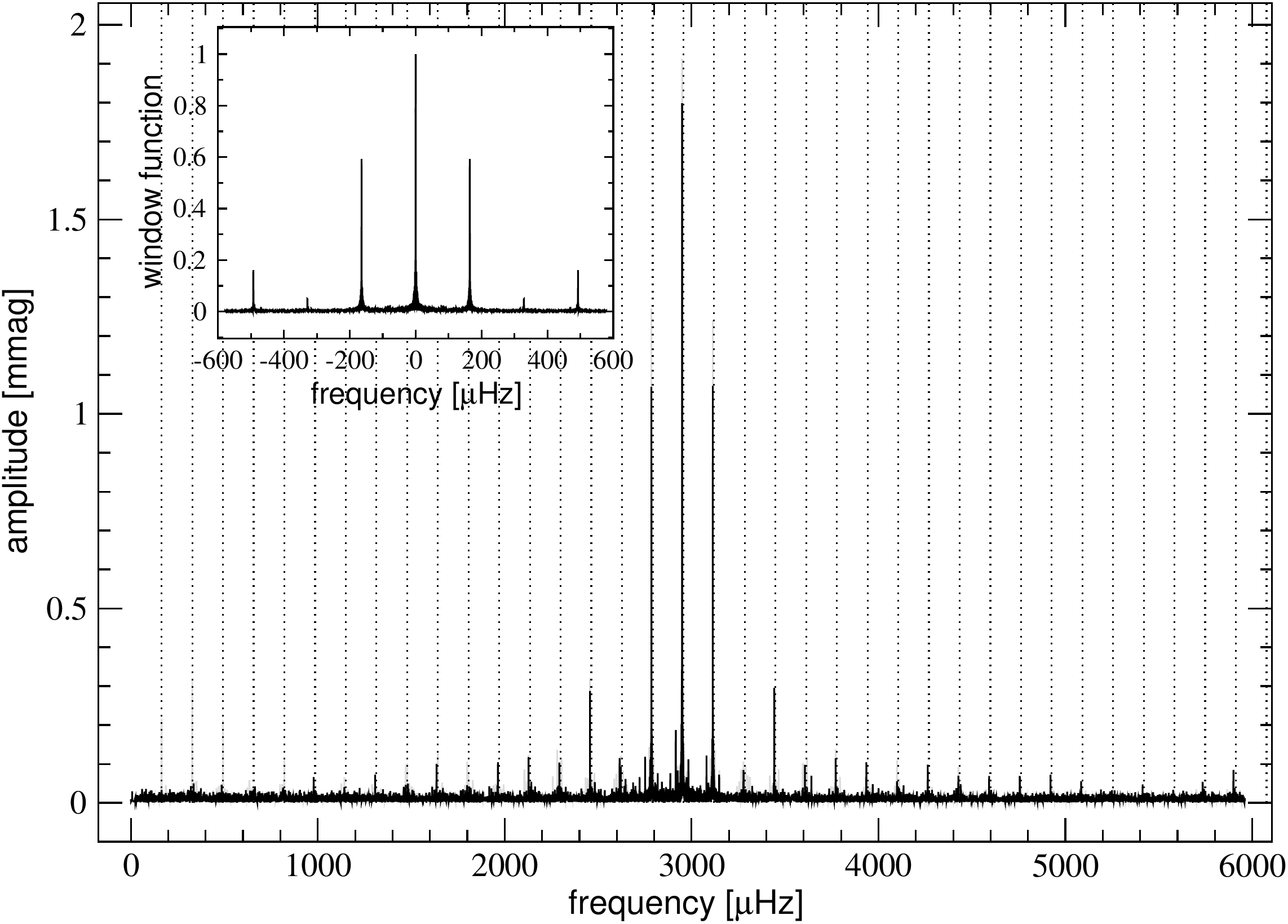}
	\caption{Same as Fig.\,\ref{fig:hd9289_ampspec} but for HD\,134214. Notice that the harmonic to the main frequency can be seen as a small peak at 5900 \muHz.}
	\label{fig:hd134214_ampspec}
\end{figure}

\begin{figure}[t]	
\centering
	\includegraphics[width=\columnwidth]{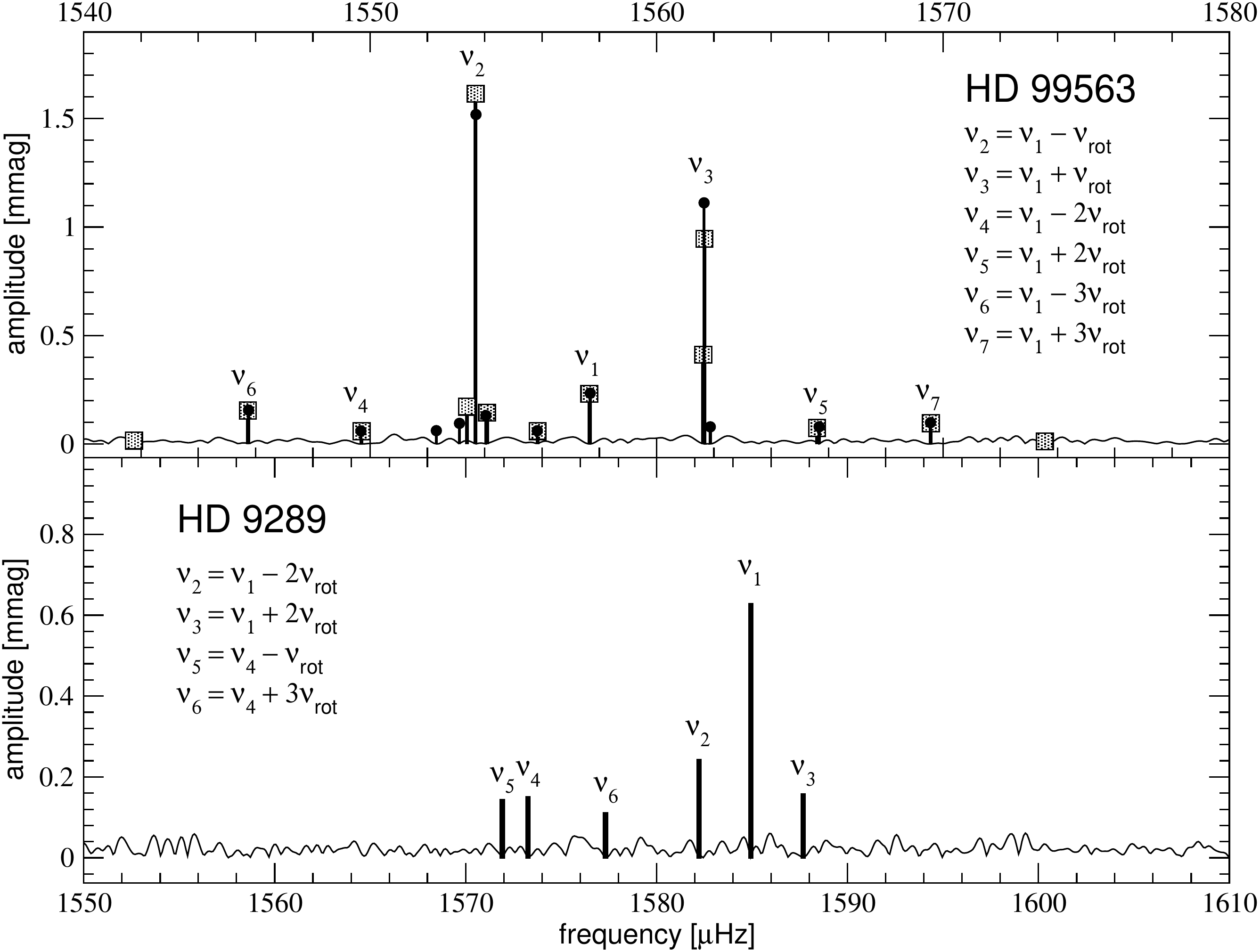}
	\caption{Schematic pulsation spectrum of HD\,99563 (top) and HD\,9289 (bottom). Labels indicate multiplet frequencies and corresponding combinations. The thin lines show the residual amplitude spectra after
	pre-whitening of the identified frequencies. The top panel shows both the unconditional (black circles) and forced (grey squares) solutions for HD\,99563. Note the different scales for both panels.}
	\label{fig:hd9xcombo}
\end{figure}

\begin{figure}[t]	
\centering
	\includegraphics[width=\columnwidth]{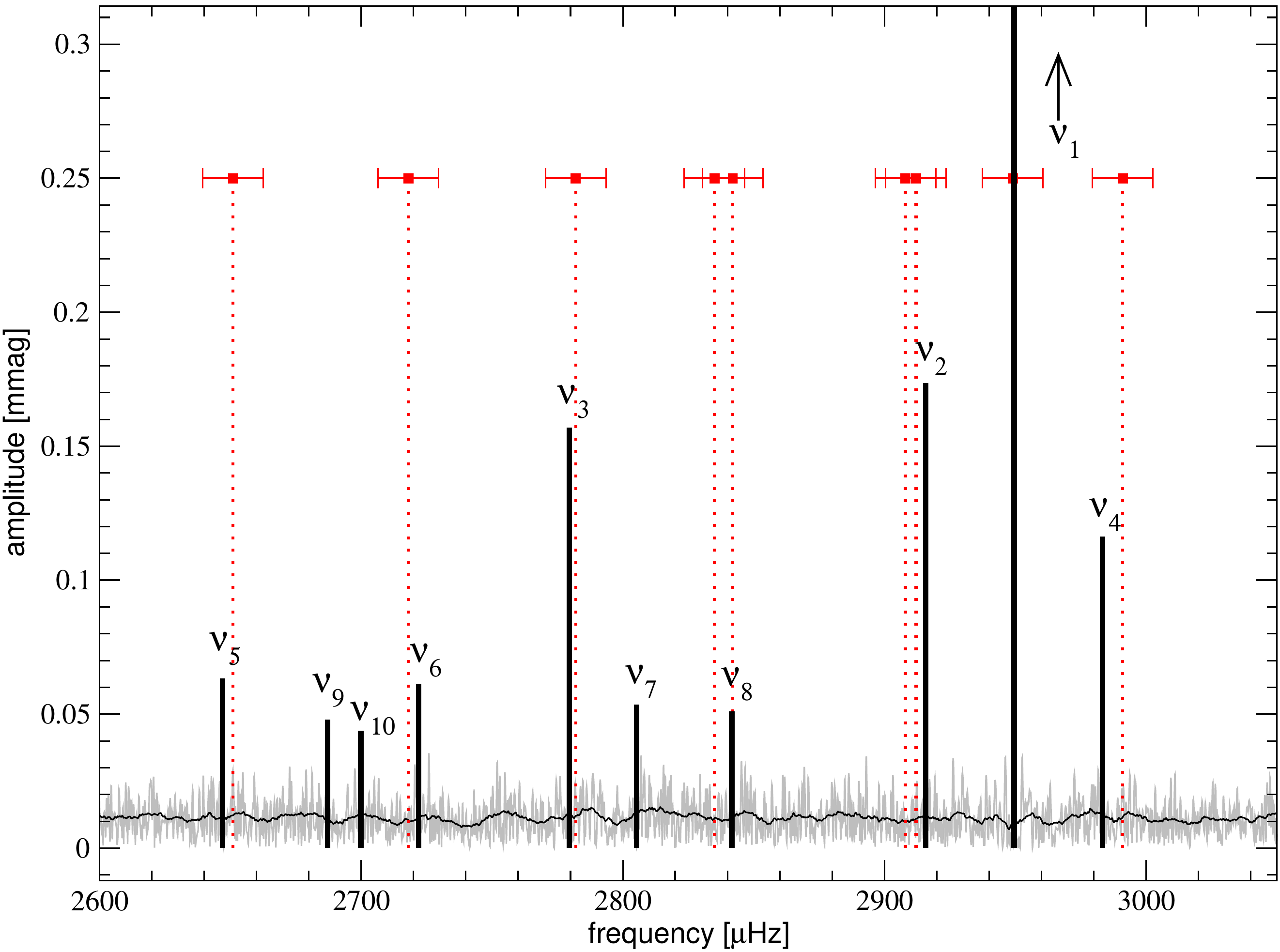}
	\caption{Schematic pulsation spectrum of HD\,134214. Squares indicate the frequencies reported in \cite{kurtz07}. The error bars show our estimated upper limits for their uncertainties derived from
	their data set length. The arrow in the upper right corner indicates that the main pulsation frequency has a much higher amplitude ($\approx 2$ mmag) which would explode the scaling of the figure. The residual amplitude spectrum after pre-whitening of the identified frequencies, and a 50-point running average, is also shown.}
	\label{fig:134214_freqana}
\end{figure}

The case of HD\,134214 is more straightforward, since we can rule out rotational modulation on timescales of our observations. As shown in Fig.\,\ref{fig:hd134214_ampspec}, the data reveal a main pulsation frequency at 2949.5 \muHz\, with a large amplitude of 1.8 mmag. Due to the strength of this signal the same instrumental modulation effect as in HD\,99563 can be observed to even higher integer multiples. To be most conservative in our analysis, we therefore excluded any other signal that is close to $\nu_{1} \pm n \nu_{\rm MOST}$ for any $n$. Naturally, we also excluded the usual integer multiples of the MOST orbital frequencies as well. Applying these criteria to the whole range considered (down to zero and up to 6000 \muHz), we retain 10 significant frequencies below 3000 \muHz. These are shown in Fig.\,\ref{fig:134214_freqana}. Their values are listed in Table\,\ref{tab:freqana134214}.

\begin{table}[t!]
	\caption{Frequency analysis results for HD\,134214.}
	\centering
	\begin{tabular}{l r r r r r r}
	\hline
	\hline
 	id & frequency & amplitude & $\theta$ & sig & SNR \\
		 &  [\muHz]  & [mmag] &  [rad]  & &  \\ 
	\hline 
$\nu_1$ & 2949.537(1) & 1.820(8) & 0.744(4) & 4291.2 & 170.1 \\ 
$2\nu_1$ & 5899.12(2) & 0.083(8) & 0.97(9) & 25.4 & 8.2 \\ 

$\nu_2$ & 2915.70(1) & 0.174(8) & 5.13(4) & 106.2 & 15.6 \\ 
$\nu_3$ & 2779.52(1) & 0.157(8) & 0.26(5) & 88.2 & 13.2 \\ 
$\nu_4$ & 2983.33(2) & 0.116(8) & 5.81(7) & 48.4 & 10.2 \\ 
$\nu_5$ & 2646.96(3) & 0.063(8) & 3.2(0.1) & 14.8 & 5.4 \\ 
$\nu_6$ & 2721.96(3) & 0.061(8) & 3.2(0.1) & 14.0 & 5.8 \\ 
$\nu_7$ & 2805.31(4) & 0.054(8) & 2.0(0.1) & 10.6 & 4.3 \\ 
$\nu_8$ & 2841.63(4) & 0.051(8) & 4.2(0.1) & 9.5 & 4.3 \\ 
$\nu_{9}$ & 2687.18(4) & 0.048(8) & 0.3(0.2) & 8.5 & 4.3 \\ 
$\nu_{10}$ & 2699.84(4) & 0.044(8) & 0.8(0.2) & 7.1 & 3.9 \\ 	
	\hline
	\hline
	\end{tabular}
	\label{tab:freqana134214}
	\tablefoot{
	Phase $\theta$ is given according to the definition $\cos\left(2\pi\nu t - \theta\right)$ with respect to the epoch HJD\,2454575.045398.}
\end{table}

\section{Discussion}						
\label{results}

\subsection{The pulsation of HD\,9289}
\label{sec:discuss9289}
The results presented here for HD\,9289 are inconsistent with the findings of \cite{kurtz94}. Their analysis yielded three frequencies, $\nu_{\rm K, 1} = 1585.06(1)$\,\muHz, $\nu_{\rm K, 2} = 1554.79(1)$\,\muHz, and $\nu_{\rm K, 3} = 1605.39(1)$\,\muHz. The authors acknowledged, however, that due to severe $1 \rm d^{-1}$ aliasing the frequencies might have been confused with their aliases. A comparison to Table\,\ref{tab:freqana9289} shows that the primary frequency is similar in both data sets but the exact values do not agree within the uncertainties. We do not find any frequencies that have values close to $\nu_{\rm K, 2}$ or $\nu_{\rm K, 3}$, and we do not expect any instrumental artefacts that would possibly complicate the detection of these frequencies. However, $\nu_{\rm K, 3}$ is consistent with an alias of $\nu_2$, with $\nu_{\rm K, 3} = \nu_2 + 2c$ within the uncertainties. Here, $c \approx 11.574$ is the conversion factor from $\rm d^{-1}$ to \muHz. A possible match for $\nu_{\rm K, 2} \approx \nu_6 - 2c$ is much more strenuous and the disagreement is too large to reconcile it with the uncertainties. In any case, it is plausible that the results from \cite{kurtz94} suffered a misidentification due to aliasing. It is not impossible, however, that the different results stem from an intrinsic change in HD\,9289's pulsation behaviour.

We know from the rotational modulation of the light curve that $v_{\rm rot} \approx 1.35$\,\muHz. Thus, the spacing in the main ``triplet" ($\nu_1$, $\nu_2$, $\nu_3$) appears to be twice the rotation frequency. Similarly, the second multiplet ($\nu_4$, $\nu_5$, $\nu_6$) also contains information about the rotation in its spacings. Whatever the nature of these modes is, if our interpretation is correct, this allows us to obtain an estimate of the rotation frequency which is independent of the mean variations induced by spots. Using $1/\sigma_{\nu}^2$ as weights, where $\sigma_{\nu}$ are the $1\sigma$ frequency uncertainties derived as described in Sect.\,\ref{sec:freqana}, we calculated the weighted average of the spacings of adjacent peaks in the multiplets. Note that this corresponds to a weighting with the spectral significance (or the square of the SNR), since $\sigma_{\nu}^2 \propto \rm 1/sig \propto \rm 1/SNR^2$ \citep{kallinger08}. The result is $\nu_{\rm rot, puls} = 1.361(7)$\,\muHz, corresponding to a period $P_{\rm rot}= 8.50(4)$\,d. This estimate is more precise than the estimate from PDM. 
Therefore, invoking the OPM, we actually observe components of (at least) a quintuplet with two components undetected ($\nu_1 + \nu_{\rm rot}$ and $\nu_1 - \nu_{\rm rot}$). Hence, under this assumption the mode cannot be described by a pure dipole. The question that remains is whether the undetected components are actually missing, or whether they are simply too small to be significant in the MOST data. 

\begin{figure}[t]	
\centering
	\includegraphics[width=\columnwidth]{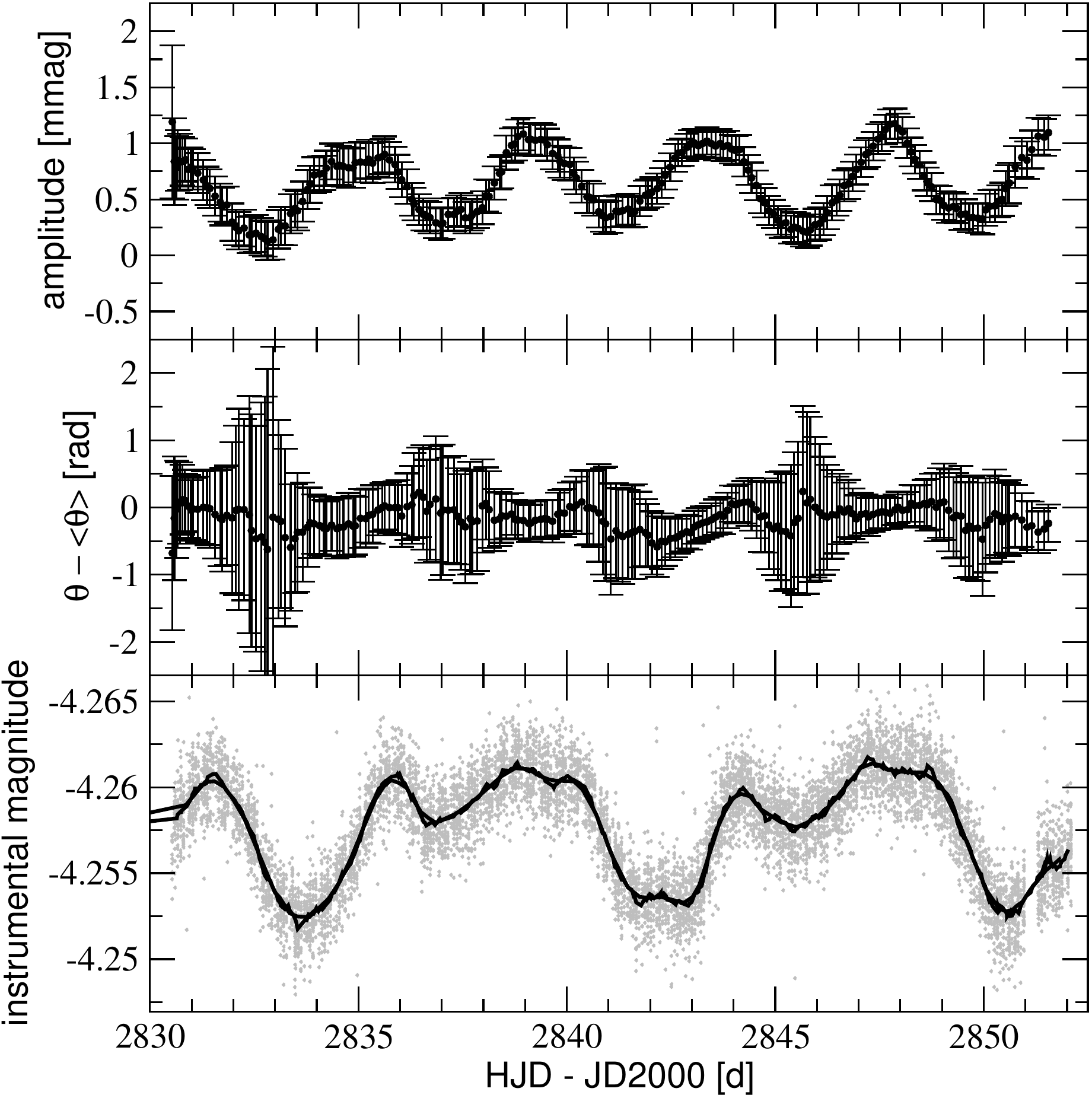}
	\caption{Time-resolved analysis of HD\,9289's main pulsation multiplet. The variation of the mean stellar brightness is shown in the bottom panel for comparison.}
	\label{fig:HD9289_tr_maintrip}
\end{figure}

Fig.\,\ref{fig:HD9289_tr_maintrip} shows a time-resolved analysis of HD\,9289's pulsation. We fitted a cosine with fixed frequency $\nu_1$ but free amplitude and phase to small subsets of the light curve after pre-whitening of all frequencies not related to the main triplet. A subset length of 1\,d, analyzed in steps of 0.1\,d, was chosen as the best compromise between reducing the uncertainties of the fitted parameters and increasing the time resolution. The result is that the pulsation amplitude is modulated twice per rotation cycle, while the pulsation phase remains roughly constant throughout. Given the large uncertainties at epochs of low pulsation amplitude, there might be changes in phase that simply cannot be reliably detected with the current data. A strong dipole component, showing two amplitude maxima per rotation cycle connected to the two pulsation poles, would produce a phase shift of $\pi$ at quadrature. The fact that we do not see a phase shift but a clear amplitude modulation, is consistent with HD\,9289 being dominated by a quadrupolar ($\ell=2$, $m=0$) pulsation mode.   

The second mode can also be explained through its spacing as a multiplet with undetected components. Assuming the simplest solution, a quintuplet, the two ``missing" components are the central component $\nu_{\rm c}$ and $\nu_{\rm c} + \nu_{\rm rot}$. Another possibility is a septuplet centred on $\nu_4$ (or alternatively $\nu_4 + 2\nu_{\rm rot}$, which we could not detect) with four components missing. In any case, while still consistent with the OPM, explaining this multiplet requires a strong departure from the usual ``distorted dipole" picture. Using $\nu_{\rm rot, puls}$ derived above as an estimated value for $\nu_{\rm rot}$, and calculating the position of the central (missing) frequency from a weighted mean based on the measured values of the detected multiplet components, the actual pulsation mode would have a frequency $\nu_{\rm c} = 1574.55(2)$\,\muHz\footnote[1]{Using $\nu_{\rm rot, PDM}$ instead, the central frequency of the second quintuplet would be $\nu_{\rm c} = 1574.48(2)$\,\muHz}.

\begin{figure}[t]	
\centering
	\includegraphics[width=\columnwidth]{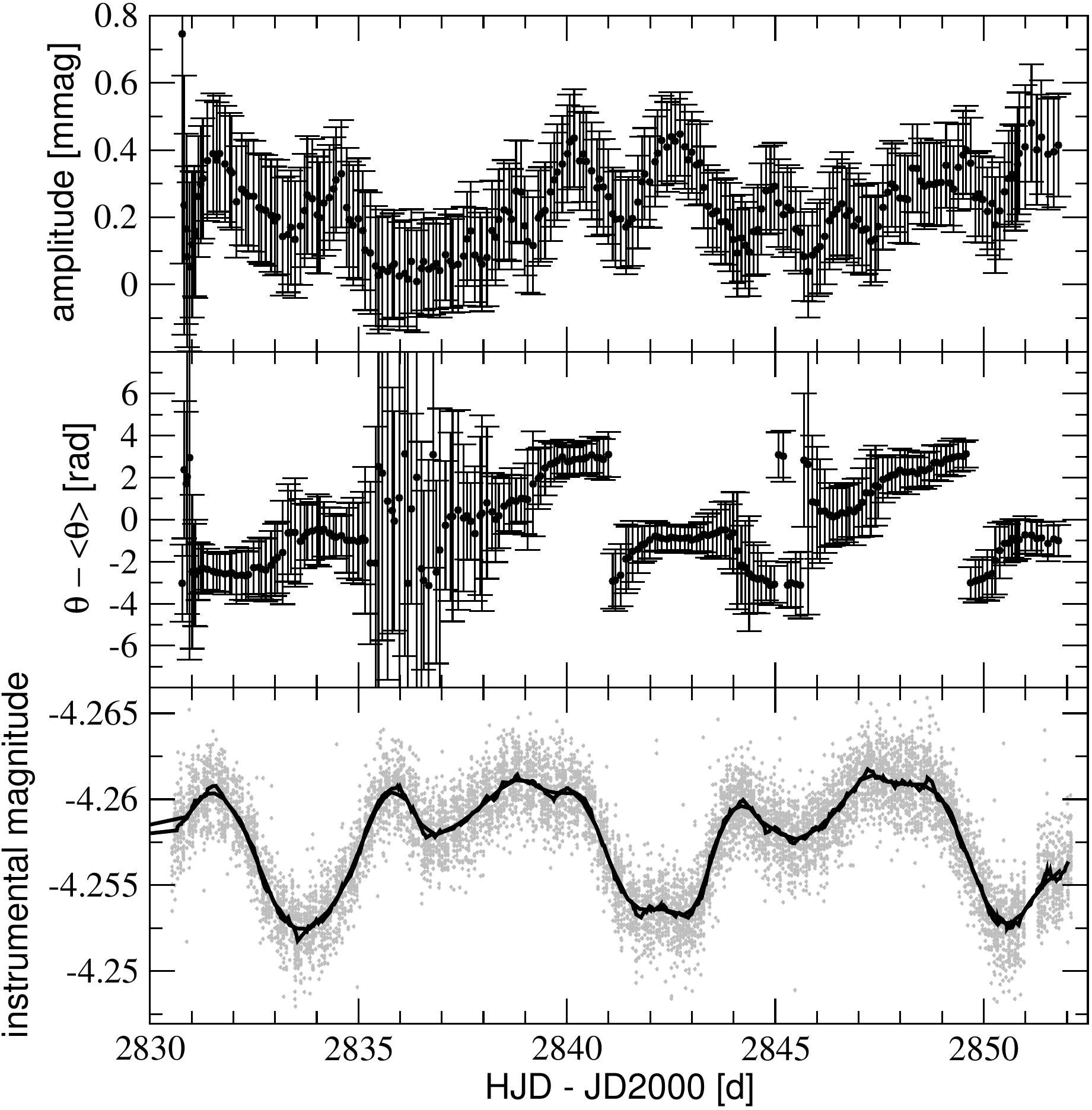}
	\caption{Time-resolved analysis of HD\,9289's secondary pulsation multiplet, assuming that the real mode has $\nu_{\rm c} = 1574.55$\,\muHz. The variation of the mean stellar brightness is shown in the bottom panel for comparison.}
	\label{fig:HD9289_tr_2ndtrip}
\end{figure}

We repeated the time-resolved analysis by fitting a cosine at fixed frequency $\nu_{\rm c}$, after subtracting all frequencies related to the main pulsation multiplet. The results are presented in Fig.\,\ref{fig:HD9289_tr_2ndtrip}. Due to the large uncertainties, the behaviour of the amplitude is not well defined, but it appears to be somewhat larger around the minimum in mean light. The phase changes continuously and linearly with time and jumps by $2\pi$ after about 1 rotation period. From the point of view of frequency analysis, this is due to the presence of $\nu_4$ and the actual absence of $\nu_{\rm c}$. The time-resolved analysis essentially tries to compensate for picking the ``wrong" frequency by shifting the phase continuously. However, if the physical mode frequency is $\nu_{\rm c}$, the continuous change in phase is real. Alternatively, we could choose $\nu_4$ as the physical mode frequency, thus requiring a septuplet with missing components. In this scenario, the phase does not show any jumps and remains rather constant while the amplitude modulation looks the same. We summarize that the SNR is too low for a conclusive time-resolved analysis of the second multiplet.

If both modes are considered quintuplets, a principle interpretation is that both modes are dominated by ($\ell = 2$, $m=0$). However, in this case they should show similar relative amplitudes (or amplitude ratios) for the various components. This is clearly not the case. Different pulsation axes might explain this observation, but we refrain from speculation at this point. Several roAp stars recently observed with the Kepler satellite show comparable multiplet structures, and strong rotational modulation due to surface features \citep[][in press]{balona_submitted}, or even give direct evidence for different pulsation axes \citep[][in press]{kurtz_submitted}. Hopefully, these more detailed observations will help us to also improve our understanding of HD\,9289.

\subsection{The pulsation of HD\,99563 - a new candidate frequency}

\begin{figure}[t]	
\centering
	\includegraphics[width=\columnwidth]{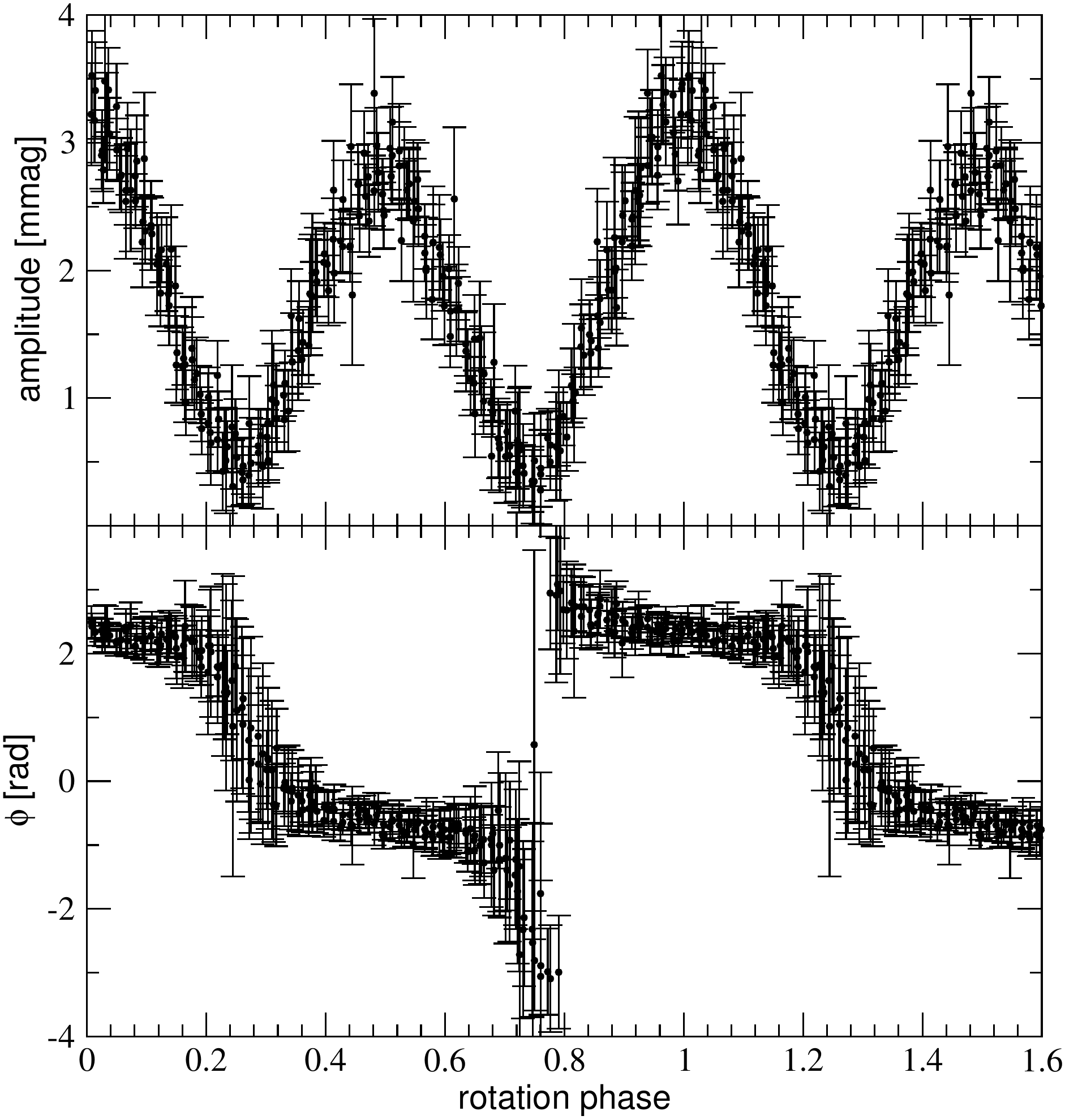}
	\caption{Time-resolved analysis of HD\,99563's main pulsation mode, choosing the frequency value displayed in Table\,\ref{tab:forcedfreqana99563}. Note that the phase $\theta$ (as in $\cos\left(\omega t - \theta\right)$) has been transformed to $\phi$ (as in $\sin\left(\omega t + \phi\right)$) for comparison to \cite{handler06}.}
	\label{fig:HD99563_tr}
\end{figure}

Fig.\,\ref{fig:HD99563_tr} shows a time-resolved analysis of the main pulsation mode of HD\,99563 after transforming the pulsation phase and defining the rotation phase so that a comparison with Fig. 6 from \cite{handler06} is feasible. As can be seen, the result is virtually indistinguishable. The difference in the two amplitude maxima is marginal, and the behaviour of the pulsation phase is exactly the same. This result does not depend on which of the two solutions for the exact frequency derived in Sect.\,\ref{sec:hd99563_freqana} we chose. Also, the harmonics and linear combinations above 3000\,\muHz\, are the same within the error bars for both solutions. 

However, there are differences in the amplitudes of the most significant multiplet components between the forced and the unconditional solution. This is due to the change in phase that comes with a change in frequency. As the relative phases of the components must change with a change in frequency, their relative amplitudes must also change in order to reproduce the light curve behaviour. Since the change in residuals between the forced and unconditional solution are negligible, there is no strong preference for either solution. However, if the values and uncertainties given in \cite{handler06} are correct, and assuming that HD\,99563 has not changed and that the OPM is applicable, the forced solution is to be preferred. Concerning the multiplet, the remaining difference between our results and those by \cite{handler06}, in particular the relative amplitudes, are expected due to the different bandpasses used the in photometric observations. Just as the amplitude of the rotational modulation varies from filter to filter, so does the amplitude modulation of the distorted dipole. The deviations from \cite{freyhammer09} are also not surprising, since the radial velocities generally show different relative amplitudes from photometry. One should be careful when using the amplitude ratios and phases to infer the general geometry of the oblique pulsator through, e.g., the technique of \cite{kurtz92}. In particular, one needs to take into account how uncertainties and instrumental properties affect the results.
 
Concerning any additional frequencies independent from the main multiplet, a number of such candidate signals were formally significant in both frequency solutions. Most of the candidates in the unconditional solution cannot be exactly reproduced by the forced solution, but instead, other significant frequencies with similar but different values emerge. The only exceptions are $\nu_{12}$ (unconditional) and $\nu_{13}$ (forced), which match in frequency, amplitude, and phase. Thus, this seems to be a signal which is independent of the exact form of the main multiplet. We therefore conclude that there is an additional candidate frequency at $\nu_{\rm cand} = 1555.84(4)$\,\muHz. It is plausible that this frequency could not be detected by \citeauthor{handler06}, since its amplitude is comparable to that of $\nu_{\rm p-2}$ and $\nu_{\rm p+2}$ which they were also unable to detect. We have no explanation for the remaining signals, formally significant but different in each solution, but we can speculate that they are artefacts caused by either an intrinsic or instrumental long-term modulation of the main pulsation frequency. This would explain why their exact values are affected by the choice of the frequency solution, while $\nu_{\rm cand}$ is unaffected. The fact that no modulation was observed by \citeauthor{handler06} increases the probability of an instrumental origin. Finally, in the forced solution, the outermost components of the nonuplet are not significant with a SNR of 1.1 and below. We can therefore not substantiate the claim of a nonuplet by \cite{freyhammer09}.

\subsection{The pulsation of HD\,134214 - a twin of HD\,24712?}

\begin{figure}[t]	
\centering
	\includegraphics[width=0.8\columnwidth]{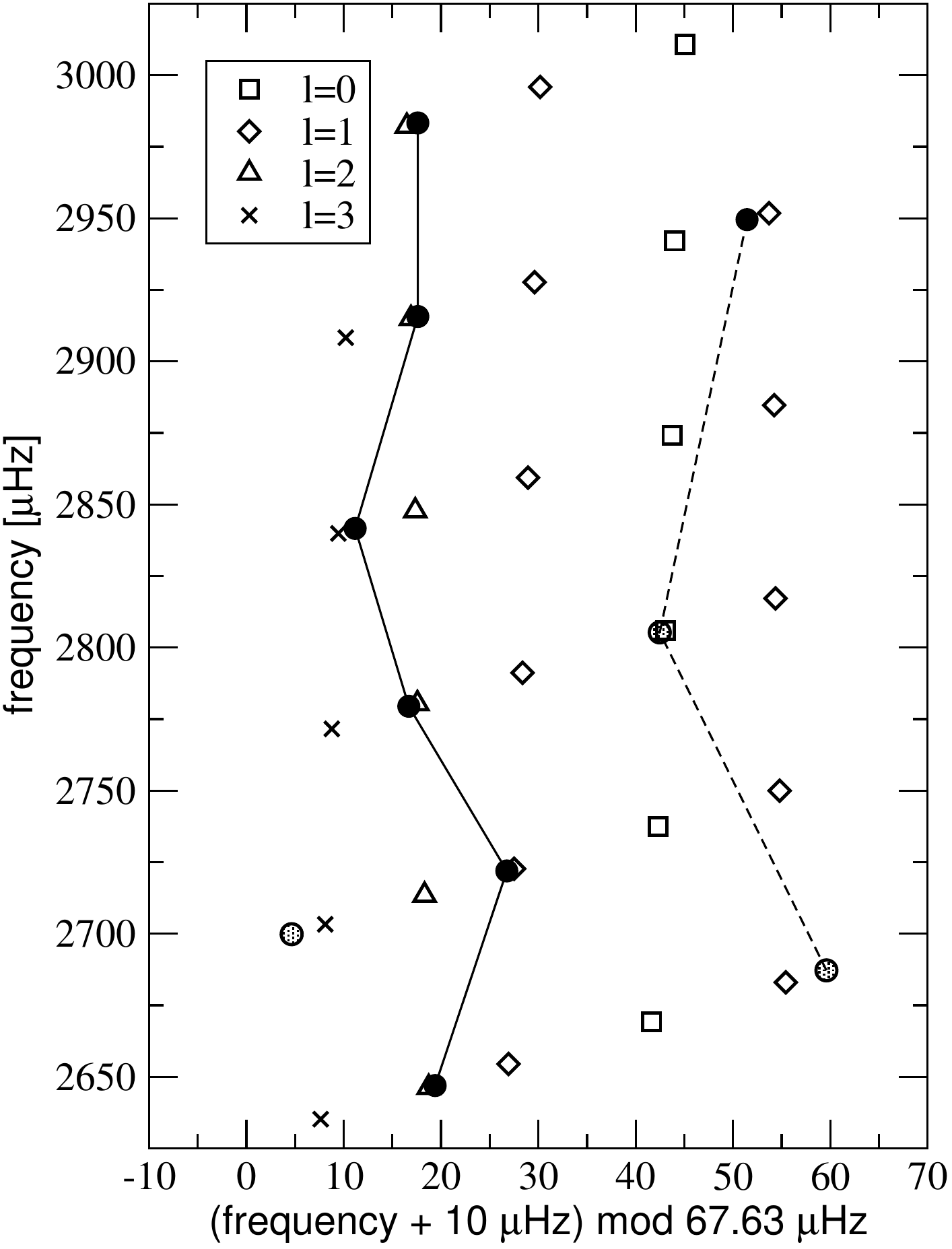}
	\caption{Echelle diagram for the frequencies extracted from the MOST data of HD\,134214. Frequencies consistent with prior observations of \cite{kurtz07} are shown as filled black circles. New candidate frequencies are shown as shaded grey circles. Other symbols correspond to a reference model (see text) with l defined as in \cite{saio2010}. The frequency uncertainties are smaller than the symbols used. Lines are visual aids to help separate the two possible mode ridges.}
	\label{fig:HD134214_echelle}
\end{figure}

As presented in Fig.\,\ref{fig:134214_freqana}, our results of a very conservative analysis for HD\,134214 show that we can confirm most frequencies reported by \cite{kurtz07}. There is a more or less regular frequency spacing that can be found throughout the region between 2600 and 3000\,\muHz. Only two of the frequencies we found do not fit this regular pattern: $\nu_3$ and $\nu_{10}$. To extract the most probable frequency spacing, we created a synthetic amplitude spectrum in which each frequency was represented by a Gaussian. The width of the Gaussians was determined using the estimated frequency uncertainties. The auto-correlation of this spectrum shows several peaks, but the most prominent peaks are a double peak at $33.80 \pm 0.05$\,\muHz,  and a single broad peak at $67.63 \pm 0.03$\,\muHz. Interpreting the latter as the large frequency separation, we created the Echelle diagram shown in Fig.\,\ref{fig:HD134214_echelle}. It immediately reveals the striking regularity in the spacing of the frequencies. Most of the frequencies line up at $\nu\,\rm mod\, 67.63\,\muHz \approx 10 \,\muHz$, but the primary frequency $\nu_1$ does not! Considering only the frequencies that are consistent with \cite{kurtz07}, $\nu_1$ is the only mode that clearly seems to relate to a different spherical degree from the remaining frequencies. Moreover, two of the new candidate frequencies seem to continue the ridge of $\nu_1$ towards lower frequencies. This strongly suggests that HD\,134214 pulsates in at least two different dominant spherical degrees. \cite{kurtz07} could not yet decide between stable and unstable frequencies, and why most of them could not be observed in ground-based photometry. However, from the point of view of the long-timebase MOST observations, the frequencies seem to be stable, and HD\,134214 reveals an impressive p-mode spectrum.

To put this result into context, the value found for the large frequency separation is very similar to that of the well-studied roAp star HD\,24712 \citep[see, e.g.,][]{saio2010}. Moreover, the frequency region of pulsation is also approximately the same. Taking the estimate of the stars' fundamental parameters from \cite{ryab07}, both are furthermore comparable in terms of magnetic field and effective temperature. However, HD\,24712 does not exhibit such large differences in amplitudes between its individual modes, shows rotational signature in its pulsation spectrum, and is found to have a higher $v \sin i$.

Putting these bits of information together allows for an interesting interpretation of the pulsation in HD\,134214. At higher magnetic field strengths the pulsation amplitude distribution for different modes can become quite complex \citep[e.g., see Fig. 12 in][]{gruberbauer08a}. This can affect our ability to detect such modes. Therefore, different dominant spherical degree components (as can be inferred from Fig.\,\ref{fig:HD134214_echelle}) and a very small inclination angle and/or obliquity (as can be inferred from the absence of rotational features and the comparison with HD\,24712) could conspire to produce the large contrast in the mode amplitudes. If the mode geometry of $\nu_1$ is strongly dipolar and concentrates most of its pulsation power very close to the rotation axis, a low inclination would result in a large observed pulsation amplitude. On the other hand, if the pulsation of the modes on the left ridge in Fig.\,\ref{fig:HD134214_echelle} is restricted to the mid-latitudes or the stellar equator, low inclinations will make those more difficult to detect. The pulsation amplitude maxima would be observed near the stellar limbs, and, due to their non-dipolar nature, they would also be subject to cancellation. This would also explain why \citeauthor{kurtz07} find the pulsation amplitudes to increase towards lines which are higher up in the atmosphere, since those are more dominant towards the stellar limbs.

As a reference, Fig.\,\ref{fig:HD134214_echelle} also shows a magnetic model produced by H. Saio (priv. comm.) using the method outlined in \cite{saio2010}. The model parameters are $M = 1.65\, M_\odot$, $\log R = 0.2694$, $\log T_{\rm eff} = 3.8584$, $\log L = 0.9250$, $X=0.70$, $Z=0.02$ and $B_{\rm p} = 4.1\,\rm kG$. Although the model does not exactly correspond to our ``two ridge" hypothesis, it supports the similarity to HD\,24712. It also identifies the main mode as being dominated by $l=1$, while most other modes are dominated by higher-degree spherical harmonics. In the case of small inclination and small obliquity, this is in support of our explanation for the large amplitude difference between $\nu_1$ and the other modes.

However, there are two problems with the picture presented above. Firstly, it cannot explain the strong amplitude contrast between $\nu_1$ and the other two frequencies on the right ridge, $\nu_7$ and $\nu_9$. The mode spectrum of HD\,24712 reveals that, over two radial orders, the amplitudes of modes of the same dominant degree can decrease by about a factor of 5. This is not enough to explain our amplitude ratios of $\sim 35$. Therefore, even invoking our hypothesis, there is still need for rather large differences in the intrinsic amplitudes and/or different dominant components. Secondly, current models suggest that for a star like HD\,24712, the amplitude of the radiative flux variation due to pulsation has its maximum at the magnetic poles, regardless of the dominant value of $l$ \citep[see Fig. 12 in][]{saio2010}. Therefore, the observed amplitudes should not have a large dependency on $l$, assuming comparable intrinsic amplitudes. If both stars are indeed very similar, and under the assumption that the models are applicable, the scenario given above is therefore unlikely to be true.

Whether or not our interpretation of HD\,134214 is correct, the situation offers exciting potential to further study the similarities and differences between HD\,134214 and HD\,24712. As a first step, asteroseismic modelling of HD\,134214 will hopefully allow us to constrain the different mode geometries from a theoretical perspective. This will be presented in an upcoming paper (Saio et al., in preparation).

\section{Conclusions}		
\label{concl}

In this paper we have presented three data sets of roAp stars obtained with the MOST satellite. All three data sets have revealed previously unknown features, or helped to clarify ambiguities that could not be resolved from the ground. 
\newline
\newline
\textbf{HD\,9289}: 
\begin{itemize}
\item We were able to derive the star's rotation period for the first time with $P_{\rm rot} = 8.50(4)$\,d.
\item At the time of observation, HD\,9289 pulsated in two modes, and showed strong amplitude variations resulting in frequency multiplets consistent with the OPM. 
\item Both modes have different multiplet structure, hinting at different mode geometries.
\end{itemize}
\textbf{HD\,99563}: 
\begin{itemize} 
\item The MOST observations present a picture mostly consistent with previous results. However, we identified a new candidate frequency independent of the primary multiplet. 
\item We were unable to confirm the presence of a nonuplet as claimed by \cite{freyhammer09}.
\end{itemize}
\textbf{HD\,134214}: 
\begin{itemize}
\item MOST reveals a rich spectrum of p modes consistent with findings by \cite{kurtz07}. Their stability was established, and the frequencies are now known to much higher precision. This turns HD\,134214 into one of the most promising roAp candidates for asteroseismology.
\item MOST also detected three new candidate frequencies, of which two fall nicely into a well-defined p-mode pattern. 
\item HD\,134214 was found to show pulsation frequencies and a large frequency spacing similar to the well-studied star HD\,24712 (HR\,1217). 
\end{itemize}

\begin{acknowledgements}
The authors dedicate this work to the memory of our dear friend and colleague Dr. Piet Reegen, who will be greatly missed by us all. The authors would like to thank Hideyuki Saio (Tohoku University) for contributing his models and valuable discussion.  
MG, DBG, AFJM, JMM, and SR acknowledge funding from the Natural Sciences \& Engineering Research Council (NSERC) Canada, and WW from the Austrian Science Funds (P 22691-N16).
\end{acknowledgements}

\bibliographystyle{aa}
\bibliography{mostroap}

\begin{thebibliography}{33}
\expandafter\ifx\csname natexlab\endcsname\relax\def\natexlab#1{#1}\fi

\bibitem[{{Balona} {et~al.}(2011){Balona}, {Cunha}, {Gruberbauer}, {Kurtz},
  {Saio}, \& {et al.}}]{balona_submitted}
{Balona}, L.~A., {Cunha}, M.~S., {Gruberbauer}, M., {et~al.} 2011, \mnras, in
  press

\bibitem[{{Bidelman} \& {MacConnell}(1973)}]{bidelman73}
{Bidelman}, W.~P. \& {MacConnell}, D.~J. 1973, \aj, 78, 687

\bibitem[{{Bigot} \& {Dziembowski}(2002)}]{bigotdziem02}
{Bigot}, L. \& {Dziembowski}, W.~A. 2002, \aap, 391, 235

\bibitem[{{Cameron} {et~al.}(2006){Cameron}, {Matthews}, {Rowe}, {Kuschnig},
  {Guenther}, {Moffat}, {Rucinski}, {Sasselov}, {Walker}, \&
  {Weiss}}]{cameron06}
{Cameron}, C., {Matthews}, J.~M., {Rowe}, J.~F., {et~al.} 2006, Communications
  in Asteroseismology, 148, 57

\bibitem[{{Cunha}(2005)}]{cunha05}
{Cunha}, M.~S. 2005, JApA, 26, 213

\bibitem[{{Dorokhova} \& {Dorokhov}(1998)}]{dorokhov98}
{Dorokhova}, T.~N. \& {Dorokhov}, N.~I. 1998, Contributions of the Astronomical
  Observatory Skalnate Pleso, 27, 338

\bibitem[{{Freyhammer} {et~al.}(2009){Freyhammer}, {Kurtz}, {Elkin}, {Mathys},
  {Savanov}, {Zima}, {Shibahashi}, \& {Sekiguchi}}]{freyhammer09}
{Freyhammer}, L.~M., {Kurtz}, D.~W., {Elkin}, V.~G., {et~al.} 2009, \mnras,
  396, 325

\bibitem[{{Gruberbauer} {et~al.}(2008){Gruberbauer}, {Saio}, {Huber},
  {Kallinger}, {Weiss}, {Guenther}, {Kuschnig}, {Matthews}, {Moffat},
  {Rucinski}, {Sasselov}, \& {Walker}}]{gruberbauer08a}
{Gruberbauer}, M., {Saio}, H., {Huber}, D., {et~al.} 2008, \aap, 480, 223

\bibitem[{{Handler} {et~al.}(2006){Handler}, {Weiss}, {Shobbrook}, {Paunzen},
  {Hempel}, {Anguma}, {Kalebwe}, {Kilkenny}, {Martinez}, {Moalusi}, {Garrido},
  \& {Medupe}}]{handler06}
{Handler}, G., {Weiss}, W.~W., {Shobbrook}, R.~R., {et~al.} 2006, \mnras, 366,
  257

\bibitem[{{Hareter} {et~al.}(2008){Hareter}, {Reegen}, {Kuschnig}, {Weiss},
  {Matthews}, {Rucinski}, {Guenther}, {Moffat}, {Sasselov}, \&
  {Walker}}]{hareter08}
{Hareter}, M., {Reegen}, P., {Kuschnig}, R., {et~al.} 2008, Communications in
  Asteroseismology, 156, 48

\bibitem[{{Huber} \& {Reegen}(2008)}]{huber08a}
{Huber}, D. \& {Reegen}, P. 2008, Communications in Asteroseismology, 152, 77

\bibitem[{{Huber} {et~al.}(2008){Huber}, {Saio}, {Gruberbauer}, {Weiss},
  {Rowe}, {Hareter}, {Kallinger}, {Reegen}, {Matthews}, {Kuschnig}, {Guenther},
  {Moffat}, {Rucinski}, {Sasselov}, \& {Walker}}]{huber08b}
{Huber}, D., {Saio}, H., {Gruberbauer}, M., {et~al.} 2008, \aap, 483, 239

\bibitem[{{Kallinger} {et~al.}(2008){Kallinger}, {Reegen}, \&
  {Weiss}}]{kallinger08}
{Kallinger}, T., {Reegen}, P., \& {Weiss}, W.~W. 2008, \aap, 481, 571

\bibitem[{{Kochukhov}(2007)}]{kochukhov07}
{Kochukhov}, O. 2007, Communications in Asteroseismology, 150, 39

\bibitem[{{Kreidl} \& {Kurtz}(1986)}]{kreidl86}
{Kreidl}, T.~J. \& {Kurtz}, D.~W. 1986, \mnras, 220, 313

\bibitem[{{Kreidl} {et~al.}(1994){Kreidl}, {Kurtz}, {Schneider}, {van Wyk},
  {Roberts}, {Marang}, \& {Birch}}]{kreidl94}
{Kreidl}, T.~J., {Kurtz}, D.~W., {Schneider}, H., {et~al.} 1994, \mnras, 270,
  115

\bibitem[{{Kurtz}(1982)}]{kurtz82}
{Kurtz}, D.~W. 1982, \mnras, 200, 807

\bibitem[{{Kurtz}(1992)}]{kurtz92}
{Kurtz}, D.~W. 1992, \mnras, 259, 701

\bibitem[{{Kurtz} {et~al.}(2011){Kurtz}, {Cunha}, {Saio}, \& {et
  al.}}]{kurtz_submitted}
{Kurtz}, D.~W., {Cunha}, M.~S., {Saio}, H., \& {et al.}, e. 2011, \mnras, in
  press

\bibitem[{{Kurtz} {et~al.}(2006){Kurtz}, {Elkin}, \& {Mathys}}]{kurtz06}
{Kurtz}, D.~W., {Elkin}, V.~G., \& {Mathys}, G. 2006, \mnras, 370, 1274

\bibitem[{{Kurtz} {et~al.}(2007){Kurtz}, {Elkin}, {Mathys}, \& {van
  Wyk}}]{kurtz07}
{Kurtz}, D.~W., {Elkin}, V.~G., {Mathys}, G., \& {van Wyk}, F. 2007, \mnras,
  381, 1301

\bibitem[{{Kurtz} \& {Martinez}(2000)}]{kurtz00}
{Kurtz}, D.~W. \& {Martinez}, P. 2000, Baltic Astronomy, 9, 253

\bibitem[{{Kurtz} {et~al.}(1994){Kurtz}, {Martinez}, \& {Tripe}}]{kurtz94}
{Kurtz}, D.~W., {Martinez}, P., \& {Tripe}, P. 1994, \mnras, 271, 421

\bibitem[{{Lenz} \& {Breger}(2005)}]{lenz05}
{Lenz}, P. \& {Breger}, M. 2005, Communications in Asteroseismology, 146, 53

\bibitem[{{Ochsenbein} {et~al.}(1981){Ochsenbein}, {Bischoff}, \&
  {Egret}}]{ochsenbein81}
{Ochsenbein}, F., {Bischoff}, M., \& {Egret}, D. 1981, \aaps, 43, 259

\bibitem[{{Reegen}(2007)}]{reegen07}
{Reegen}, P. 2007, \aap, 467, 1353

\bibitem[{{Reegen} {et~al.}(2006){Reegen}, {Kallinger}, {Frast}, {Gruberbauer},
  {Huber}, {Matthews}, {Punz}, {Schraml}, {Weiss}, {Kuschnig}, {Moffat},
  {Walker}, {Guenther}, {Rucinski}, \& {Sasselov}}]{reegen06}
{Reegen}, P., {Kallinger}, T., {Frast}, D., {et~al.} 2006, \mnras, 367, 1417

\bibitem[{{Rowe} {et~al.}(2006){Rowe}, {Matthews}, {Kuschnig}, {Guenther},
  {Moffat}, {Rucinski}, {Sasselov}, {Walker}, \& {Weiss}}]{rowe06}
{Rowe}, J.~F., {Matthews}, J.~M., {Kuschnig}, R., {et~al.} 2006, \memsai, 77,
  282

\bibitem[{{Ryabchikova} {et~al.}(2007){Ryabchikova}, {Sachkov}, {Kochukhov}, \&
  {Lyashko}}]{ryab07}
{Ryabchikova}, T., {Sachkov}, M., {Kochukhov}, O., \& {Lyashko}, D. 2007, \aap,
  473, 907

\bibitem[{{Saio} {et~al.}(2010){Saio}, {Ryabchikova}, \& {Sachkov}}]{saio2010}
{Saio}, H., {Ryabchikova}, T., \& {Sachkov}, M. 2010, \mnras, 403, 1729

\bibitem[{{Shibahashi}(2003)}]{shiba03}
{Shibahashi}, H. 2003, in Astronomical Society of the Pacific Conference
  Series, Vol. 305, Astronomical Society of the Pacific Conference Series, ed.
  L.~A. {Balona}, H.~F. {Henrichs}, \& R.~{Medupe}, 55

\bibitem[{{Stellingwerf}(1978)}]{stellingwerf78}
{Stellingwerf}, R.~F. 1978, \apj, 224, 953

\bibitem[{{Walker} {et~al.}(2003){Walker}, {Matthews}, {Kuschnig}, {Johnson},
  {Rucinski}, {Pazder}, {Burley}, {Walker}, {Skaret}, {Zee}, {Grocott},
  {Carroll}, {Sinclair}, {Sturgeon}, \& {Harron}}]{walker03}
{Walker}, G., {Matthews}, J., {Kuschnig}, R., {et~al.} 2003, \pasp, 115, 1023

\end{thebibliography}

\end{document}